\documentclass[12pt]{article}
\usepackage[centertags]{amsmath}
\usepackage{amssymb}
\usepackage{graphicx}
\usepackage{mathrsfs}
\usepackage{epsfig}
\usepackage{graphicx}
\usepackage{amssymb}
\usepackage{amsmath}
\usepackage{epsfig}
\usepackage{dcolumn}
\usepackage{rotating}
\usepackage{color}

\newcommand{\be}{\begin{equation}}
\newcommand{\ee}{\end{equation}}
\newcommand{\bea}{\begin{eqnarray}}
\newcommand{\eea}{\end{eqnarray}}

\begin{document}

\title{\textbf{Holographic estimation of multiplicity and membranes collision in  modified spaces  $\mathrm{AdS}_5$}}

\author {I.Ya. Aref'eva$^{1}$\footnote{E-mail: arefeva@mi.ras.ru} ,
E.O. Pozdeeva$^{2}$\footnote{E-mail: pozdeeva@www-hep.sinp.msu.ru} , T.O.
Pozdeeva$^{3}$\footnote{E-mail: pozdeeva@inbox.ru}  \vspace*{3mm} \\
\small $^1$ Steklov Mathematical Institute of Russian Academy of Science, Moscow, Russia \\
\small  $^2$ Skobeltsyn Institute of Nuclear Physics,  Lomonosov Moscow State University, Moscow, Russia\\
\small  $^3$ Moscow Aviation Institute (National Research University), Moscow, Russia }
\date{ }

\maketitle

\begin{abstract}
The quark-gluon plasma formed as a result of heavy-ion collisions is currently investigated actively both
theoretically and experimentally. According to the holographic approach, forming a quark-gluon plasma
in the four-dimensional world is associated with creating black holes in a five-dimensional anti-de Sitter
space. The multiplicity of particles produced in heavy-ion collisions is then determined by the
entropy of the five-dimensional black hole, which is estimated by the area of the trapped surface.
To fit the experimental data for multiplicity Kiritsis and Taliotis
have proposed to consider black holes formation in modified $\mathrm{AdS}_5$ spaces with different $b$-factors. In this paper we  consider the formation of
black holes under collision of membranes in modified $\mathrm{AdS}_5$ spaces with $b$-factors.
Following the previous proposals we consider the power-law and exponential
 $b$-factors, as well as mixed types of $b$-factors.
 We study dynamics of
the change of the trapped surface area depending on the energy for each investigated space. We find that the power-law and
mixed factors fit better to the experimental date.

\end{abstract}

\section{Introduction}\label{sec1}

The AdS/CFT duality is a powerful method for studying quantum systems in situations where the
ordinary perturbation theory is inapplicable \cite{1}--\cite{3}, for the recent review see \cite{CasalderreySolana:2011us}. The description of the quark-gluon plasma (QGP)
formation in heavy-ion collisions using the idea of AdS/CFT duality has recently been actively developed, see \cite{4} and refs there in.
The QGP formation process (thermalization) is then interpreted as a black hole formation process in an
auxiliary five-dimensional anti-de Sitter ($\mathrm{AdS}_5$) space. The formation of black holes in the $\mathrm{AdS}_5$ space is
considered both using the analysis of shock waves \cite{5}--\cite{12} and using the Vaidya metric (see \cite{14}--\cite{16} and
the references therein). Such a method allows deriving the physical characteristics of the quantum four-dimensional
system based on results obtained in the $\mathrm{AdS}_5$ space for a classical system. In particular, one can get the
holographic estimation of the multiplicity of particle production in heavy-ion collisions.
The multiplicity is assumed to be determined by the entropy of the black
hole created in the $\mathrm{AdS}_5$ space. This hypothesis allows estimating the dependence of the multiplicity on
the energy and comparing it with the experimental data already obtained \cite{17}.

The elementary dual models considered in \cite{5}--\cite{ABP} require modification \cite{12} to describe the experimental
data more precisely. The problem of black hole formation for modified models has been considered  by Kiritsis and Taliotis in the
case of point-like sources \cite{12}.
 However these calculations are rather complicated, and it is therefore interesting to use domain walls as a model of
colliding ions. This approach was proposed in \cite{9}. A set of problems associated with the infinite sizes of
the domain walls then arises, but the regularization in which a finite wall size is introduced, as we showed
in \cite{ABP}, allows using this method. Using domain walls significantly simplifies the problem
for modified models.

In this paper we consider domain-walls (membranes) collision  in a modified $\mathrm{AdS}_5$ space. We follow the modifications proposed in \cite{Gursoy:2007cb}-\cite{Kiritsis 2009}. The basis for the proposed modifications is
the introduction of $b$-factors of the power-law, exponential, and mixed types. Our main goal here is to
estimate the dependence of the trapped surface area on the energy of colliding walls for different types of $b$-factors and
 use this estimation to get  dependence of the multiplicity on the energy of colliding ions.

The paper is organized as follows. In Sect. \ref{sec2} we remind the main facts about the $b$-factor modification of $\mathrm{AdS}_5$.
 Here we consider power-law, exponential and mixed types $b$-factor.
 In Sect. \ref{sec2.1} we pay a special attention to potentials of the scalar field responsible for the given $b$-factors. Then in Sect. \ref{ssec2.2} and Sect. \ref{ssec2.3}   we present the shock waves for point-like sources as well as for domain walls in these modified $\mathrm{AdS}_5$ spaces.   In the end of our set-up we present equations defined the trapped surface.
  In Sect. \ref{sec3} we perform  as estimations of the area of the trapped surface produce in collisions  of two shock domain walls in the modified
 $\mathrm{AdS}_5$ with the power-law $b$-factor. In Sect. \ref{sec4} we do the same for the exponential $b$-factor. In Sect. \ref{sec5} and Sect. \ref{sec6} we consider the mixed $b$-factors.

\section{Set up}\label{sec2}
\subsection{Background} \label{sec2.1}
We consider the action of five-dimensional gravity coupled to a scalar dilaton field  in the presence of a negative cosmological constant
\begin{equation}
\label{ex1}
 S_5=S_{R}+S_{\Phi},
\end{equation}
here, $S_{R}$ is the Einstein-Hilbert action with the negative cosmological constant
\begin{equation*}
 S_{R}=-\frac{1}{16\pi G_5}\int\sqrt{-g}\,\biggl[R+\frac{d(d-1)}{L^2}\biggr]dx^5,
\end{equation*}
$d+1=D=5$, $S_{\phi}$ is the dilaton action,
\begin{equation*}
 S_{\Phi}=-\frac{1}{16\pi G_5}\int\sqrt{-g}\,\biggl[-\frac{4}{3}(\partial\Phi)^2+V(\Phi)\biggr]dx^5 \label{Dilaton Action}.
\end{equation*}
It is assumed that the background metric has the form
\begin{equation}
\label{ex2}
 ds^2=b^2(z)\bigl(dz^2+dx^i\,dx^i-dx^{+}\,dx^{-}\bigr),\qquad i=1,2,
\end{equation}
In this paper we consider several  types of $b$-factors \cite{12}.
The Einstein equations reduce to two  independent relations  \cite{Gursoy:2007cb,Gursoy:2008bu}.
\begin{equation}
\label{eq5}
 \frac{3b^{\prime\prime}}{b}+\frac{2}{3}(\Phi')^2-\frac{b^2}{2}V(\Phi)-\frac{6b^2}{L^2}=0,
\end{equation}
\begin{equation}
\label{eq6}
 \frac{6(b^\prime)^2}{b^2}-\frac{2}{3}(\Phi')^2-\frac{b^2}{2}V(\Phi)-\frac{6b^2}{L^2}=0,
\end{equation}
where $b=b(z),$ $b^\prime=\partial _z b$
and  the dilaton field depends only on  $z$, $\Phi=\Phi(z)$.

 The scalar field equation is
\begin{equation}\label{seq}
  \frac{1}{b^5}\frac{\partial}{\partial z}\biggl(b^{3}\frac{\partial}{\partial z}\biggr)\Phi+\frac{3}{8}\frac{ \partial V(\Phi_{s})}{\partial \Phi_s}=0.
\end{equation}
One can see from  \eqref{eq5} and \eqref{eq6} that the  dilaton field and its potential are related to the  $b$-factor:
\begin{align}
&\Phi'=\pm\frac{3}{2}\sqrt{\biggl(\frac{2(b^\prime)^2}{b^2}-\frac{b^{\prime\prime}}{b}\biggr)}
\label{eq9},\\
& V(\Phi(z))=\frac{3}{b^2}\biggl(\frac{b^{\prime\prime}}{b}+\frac{2(b^\prime)^2}{b^2}-\frac{4b^2}{L^2}\biggr).
\label{eq8}
\end{align}
Note also that equations  \eqref{eq8} and \eqref{eq9} guarantee the validity of the field equation \eqref{seq}.

These equations for the given $b\neq b(z)$ provide a non-explicit form of potential $V=V(\Phi)$. The explicit form $V=V(\Phi)$
can be found by the superpotential method \cite{Kiritsis 2009}. To use this method it is convenient to rewrite \eqref{eq6} and \eqref{seq} in the domain wall
coordinates\footnote{$u$ is an analogue of the cosmic coordinate in cosmological application of the superpotential method, see  \cite{AKV}.}
\begin{equation}\label{u}
  u=\int b(z)dz
\end{equation}
and metric has the form
\begin{equation}\label{metric}
  ds^{2}=du^{2}+e^{2A(u)}\left( dx^{i}dx^{i}-dx^{+}dx^{+}\right),
\end{equation}
where $A(u)=\ln(b(u))$. The potential is related with the superpotential $W$ as
\begin{equation}
  V(\Phi)=-\frac43\left(\frac{dW}{d\Phi}\right)^2+\frac{64}{27}W^2-\frac{12}{L^2}.
\end{equation}
The superpotential is defined from equation the following
\begin{equation}
   W(\Phi)=-\frac{9}{4}\frac{dA}{du}
 \end{equation}
 and the dilaton field is given by
 \begin{equation}\label{Phi}
   \Phi=\pm\frac{3}{2}\int \sqrt{-\frac{d^2A}{du^2}}du.
 \end{equation}

\subsubsection{Exponential $b$-factor}
We begin by a space with $b$-factor of the form
$b={\mathrm e}^{-z/R}.$ It is assumed that $R\sim\Lambda^{-1}_{\text{\tiny\rm QCD}}\sim1\,\text{fm}.$

With the help of superpotentials method  one gets the dependence of the potential from the field
\begin{eqnarray}
 V(\Phi)=-\frac{12}{L^2}+\frac{9}{R^2}\exp\left(\pm\frac{4(\Phi_s-\Phi_0)}{3}\right),\quad \mbox{for} \quad b={\mathrm e}^{-z/R},
\end{eqnarray}
where $\Phi_{0}$ is a constant.

\subsubsection{Power-law $b$-factor}
Now we consider a space with power-law $b$-factor of the form $b(z)=\left(\displaystyle\frac{L}{z}\right)^a$. If $a=1$ we have $\mathrm{AdS}_5$ space.

The  potential and fields can be represented explicitly  thought variable $z$.
 Since in this case $\Phi=\Phi(z)$ is single-valued function we can find $z=z(\Phi)$ and
 substitute it to the  expression for potential $V(z)$ to get
\begin{equation}
  V(\Phi)=-\frac{12}{L^2}+\frac{3a(3a+1)}{{L}^{2a}}\exp{\left(\pm\frac{4}{3}\sqrt{\frac{(a-1)}{a}}(\Phi-\Phi_0)\right)}.
\end{equation}
 $V(\Phi)$ is real for $a>1$.

To give a meaning to $b(z)=\biggl(\displaystyle\frac{L}{z} \biggl)^{a}$ with $a<1$ one can consider the phantom field $\Phi_{p}$ with the action
\begin{equation*}
 S_{\Phi_{p}}=-\frac{1}{16\pi G_5}\int\sqrt{-g}\,\biggl[\frac{4}{3}(\partial \Phi_{p})^2+\tilde{V}(\Phi_{p})\biggr]dx^5.
\end{equation*}
The phantom field is related with the  dilaton field $\Phi$ via
  $\Phi-\Phi_0=i(\Phi_{p}-\Phi_{p_{0}})$, and the potential for $a<1$ becomes
     $$ \tilde{V}(\Phi_{p})=-\frac{12}{L^2}+ \frac{3a(3a+1)}{{L}^{2a}}\exp\left(\pm\frac43\sqrt{\frac{1-a}{a}}(\Phi_{p}-\Phi_{p_0})\right).$$
  Instead of relations (\ref{eq9}),(\ref{eq8}) for the phantom field one gets
 \begin{align}
&\partial_{z} \Phi_{p} =\pm\frac{3}{2}\sqrt{\biggl(\frac{b^{\prime\prime}}{b}-\frac{2(b^\prime)^2}{b^2}\biggr)},\quad  \tilde{V}(\Phi_{p}(z))=\frac{3}{b^2}\biggl(\frac{b^{\prime\prime}}{b}+\frac{2(b^\prime)^2}{b^2}-\frac{4b^2}{L^2}\biggr).
\end{align}

\subsubsection{Mixed factor of the form $b(z)=\left(\displaystyle\frac{L}{z}\right)^a \exp\left(-z^2/R^2\right)$}

Now we consider a space with the mixed $b$-factor of the form
 \be
\label{mixed}
b(z)=\left(\frac{L}{z}\right)^a \exp\left(-z^2/R^2\right).
 \ee
 The superpotentials method cannot be applied to this case,
since  $z$-variable cannot be represented through a new  $u$-variable explicitly and we cannot represent $b(u)$  explicitly.
 For this  $b$-factor we can express  $\partial_z\Phi(z)$ and $V(\Phi(z))$ using \eqref{eq9}, \eqref{eq8} as

 \begin{eqnarray}\label{com}
                      && \partial_z\Phi(z)=\pm\frac{3}{2}\frac{\xi}{R^2\,z}   ,\,\mbox{where}\, \quad \xi=\sqrt{\zeta}, \quad \zeta=4z^4+2R^2(2a+1)z^2+aR^4(a-1) , \\
            &&V(z)=-\frac{12}{L^2}+\frac{3\left(\frac{L}{z}\right)^{-2a}\left(aR^4(3a+1)+2z^2R^2(6a-1)+12z^4\right)\exp{\left(\frac{2z^2}{R^2}\right)}}{z^2R^4}. \end{eqnarray}
Integrating \eqref{com} we get
\begin{equation}\Phi_\pm=\pm\left(\frac34\frac{\xi}{R^2}+\frac{3}{8}\left(2a+1\right)\ln\left(\xi+\frac{(2a+1)R^2+4z^2}{2}\right)\right.-\label{PhiBigA}\end{equation}
$$\left.\frac34\sqrt{a(a-1)}\ln\left(\frac{2R^2\left\{a(a-1)R^2+(2a+1)z^2+\xi\sqrt{a(a-1)}\right\}}{z^2}\right)\right)+\Phi_{0\pm}.$$

For $a>1$ all expressions in \eqref{PhiBigA} are well defined.  We can find $V(\Phi)$  at $z\rightarrow\infty,$ $\Phi\sim\frac{3}{2}\frac{z^2}{R^2}$ and $V\sim\Phi^{a+1}\, e^{\frac{4}{3}\Phi}.$

\begin{figure}[h!] \centering
\includegraphics[height=4cm]{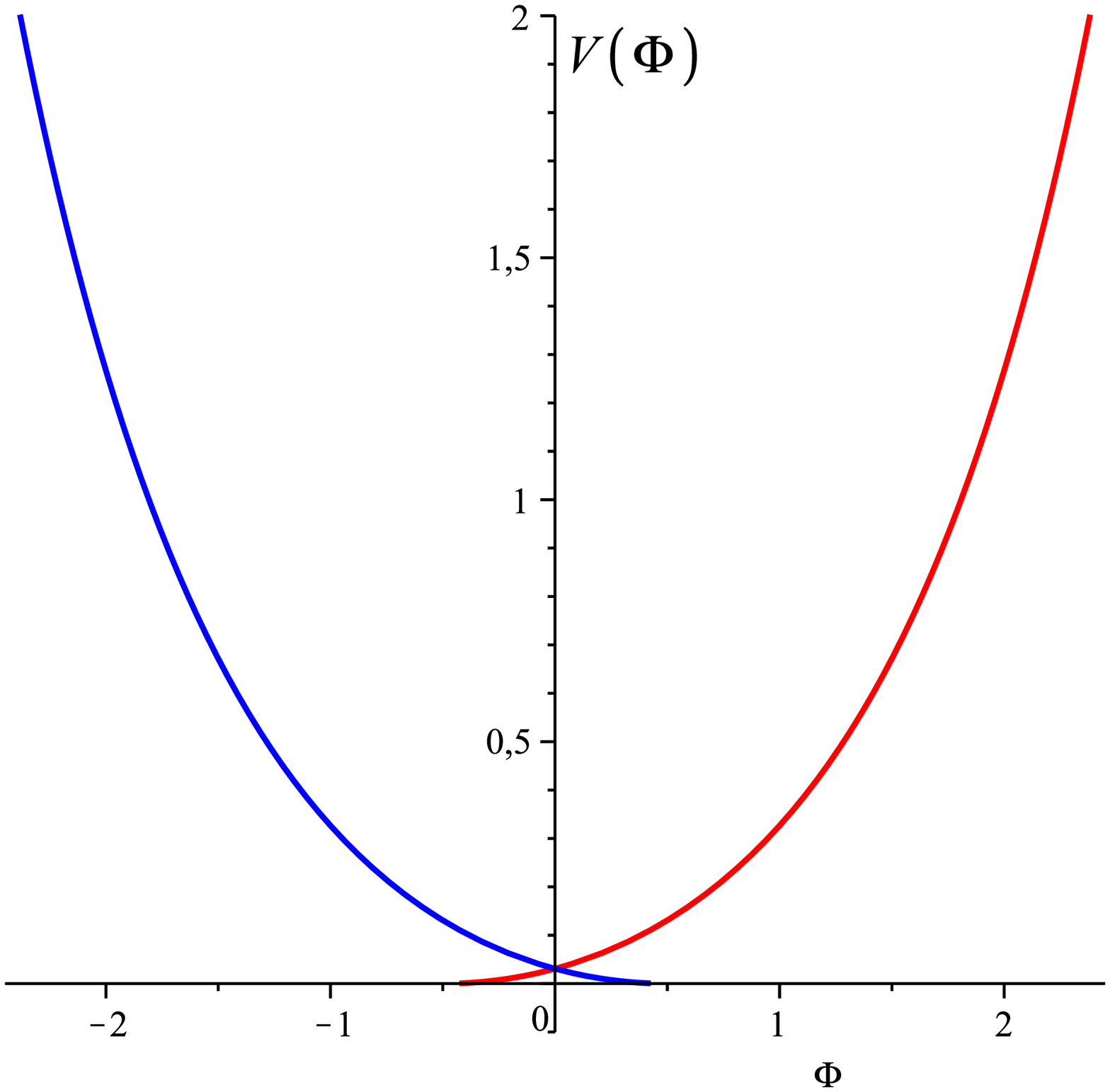} A.\,\,\,\qquad\qquad
\includegraphics[height=4cm]{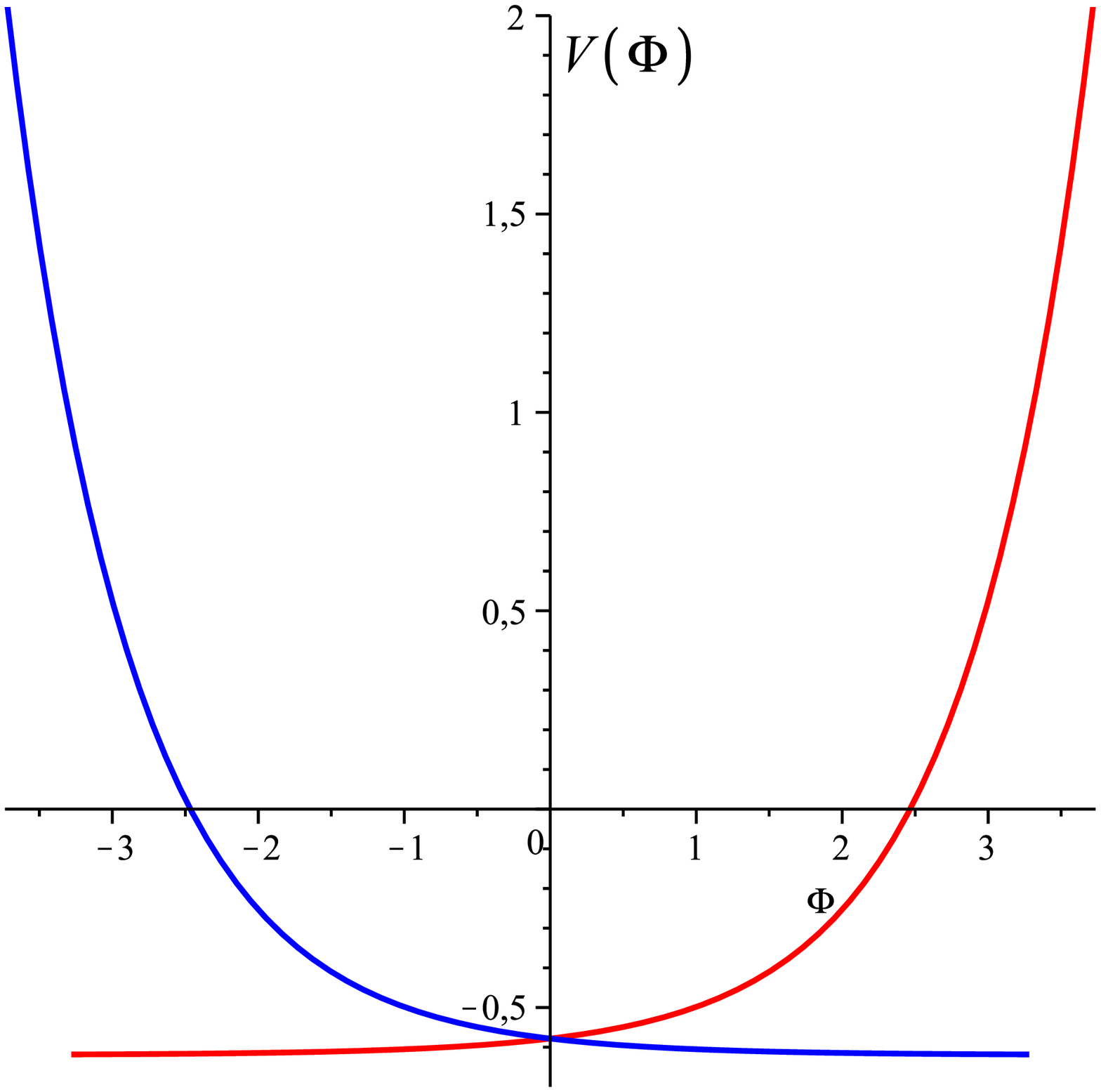} B.
\caption{ The potentials  corresponding to $b(z)$ in the form (\ref{mixed}) at $a=2,$ $L=4.4$ fm, $R=1$ fm
(the left panel) and  at $a=1,$ $L=4.4$ fm, $R=1$ fm (the right panel) and
 $\Phi_0 =-0.5.$ Two branches,  red  and blue, correspond to $\pm$ in \eqref{com} (the red for $+$, the blue for $-$).
 }\label{V(F)mixedA2}
\end{figure}

We  present  $V(\Phi)$ graphically in Fig \ref{V(F)mixedA2} for $a=1, \,\, a=2$.

\begin{figure}[h!] \centering
\includegraphics[height=4cm]{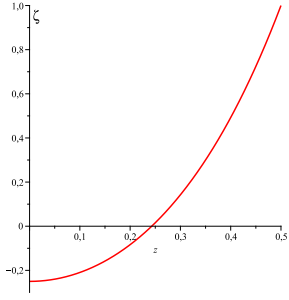}
\caption{ The function $\zeta=\zeta(z)$ at $a=1/2,$ $R=1$ fm.}\label{zeta(a=05)}
\end{figure}

For $a<1$,
 $\zeta(z)$ has a positive root, $\zeta(z_0)=0$, and $\zeta(z)>0$ for $z>z_0$, $\zeta(z)<0$ for $0<z<z_0$, see Fig \ref{zeta(a=05)}.  For example, in the case of $a=1/3$,  the point $z_0\approx0.249$ fm
 and for  $a=1/2$,  the point $z_0\approx0.243$ fm.

 Near $z=z_0$ the function $\Phi(z)$ has a singularity of the form
 \be
 \Phi(z)\sim (z^2-z_0^2)^{3/2}(A+{\cal {O}}(z^2-z_0^2)),
 \,\,\,\,\,\,\,A=\pm\frac{2(8a+1)^{1/4}}{\sqrt{z_0}},
 \ee
 $\Phi(z)$ is real for $z>z_0$,

$$\Phi_{\pm}=\frac{3}{4R^2}\left(\sqrt{\zeta}-\sqrt{-\left( {a}^{2}-a \right) {R}^{4}}
\arctan\left({\frac {\left( {a}^{2}-a \right) {R}^{4}+\left( 2\,a+1 \right) {R}^{2}{z}^{2}}{\sqrt{\zeta}\sqrt {- \left( {a
}^{2}-a \right) {R}^{4}}}} \right)+\right.$$ $$\left.\frac12\, \left( 2\,a+1 \right) {R}^{2
}\ln  \left( a \left( 4\sqrt{\zeta}+8\,{z}^{2}+2\, \left( 2\,a+1
 \right) {R}^{2} \right)  \right)  \right)+\Phi_{0\pm}, \qquad z>z_0$$
 and  becomes imaginary for $z<z_0$, $(\Phi_\pm-\Phi_{0\pm})=i(\Phi_{p\pm}-\Phi_{p0\pm})$,
 $$\Phi_{p\pm}=\pm\frac{3}{4R^2}\, \left(\sqrt {a{R}^{4} \left(1-a\right) } \ln  \left( {\frac {2\,\sqrt
{a{R}^{4} \left(1-a\right) }\sqrt { -\zeta}+2\,{R}^{2}
 \left[  \left( a-{a}^{2}\right) {R}^{2}-2\,{z}^{2} \left( a+1/2
 \right)  \right] }{{z}^{2}}} \right)\right. $$
 $$\left.- \left(  \left( a+\frac12 \right) {R}^{2}\arctan \left({
\frac { \left( 2\,a+1 \right) {R}^{2}+4\,{z}^{2}}{2\sqrt {-\zeta}}} \right) -\sqrt {-\zeta} \right)  \right)+\Phi_{0\pm},\quad z<z_0.$$
The convenient  choice of constants is $\Phi_{s\pm}(z_0)=\Phi_{p\pm}(z_0)=0$. The imaginary scalar field corresponds to the phantom
 sign of the kinetic term and one can write
 $\Phi=\Phi_s\Theta(z-z_0)+i\Phi_p\Theta(z_0-z)$ and interpret this model as a model with an alternating sign  of the kinetic term.

\begin{figure}[ht] \centering
\includegraphics[height=4cm]{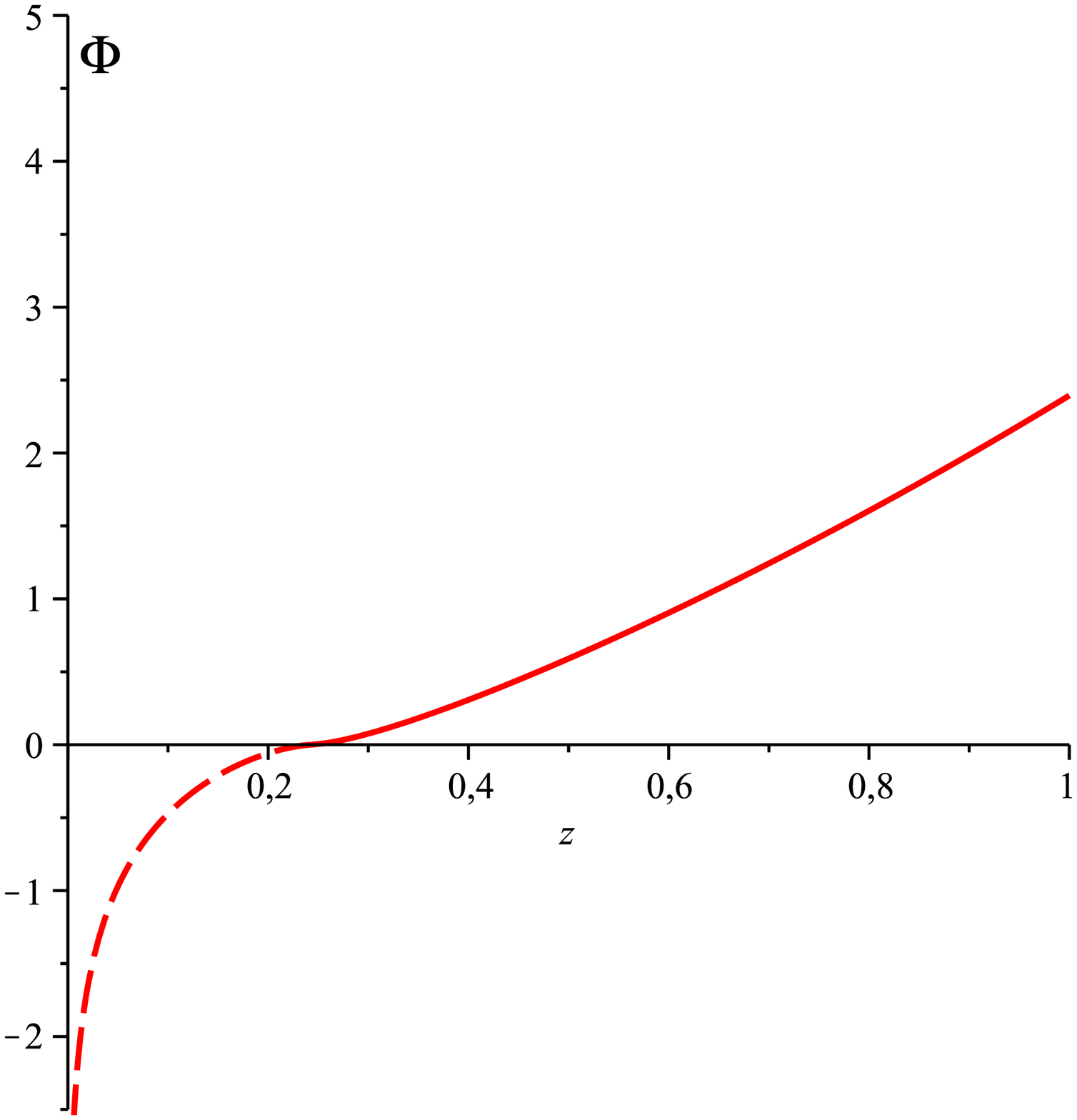}\,A.\,
\qquad\includegraphics[height=4cm]{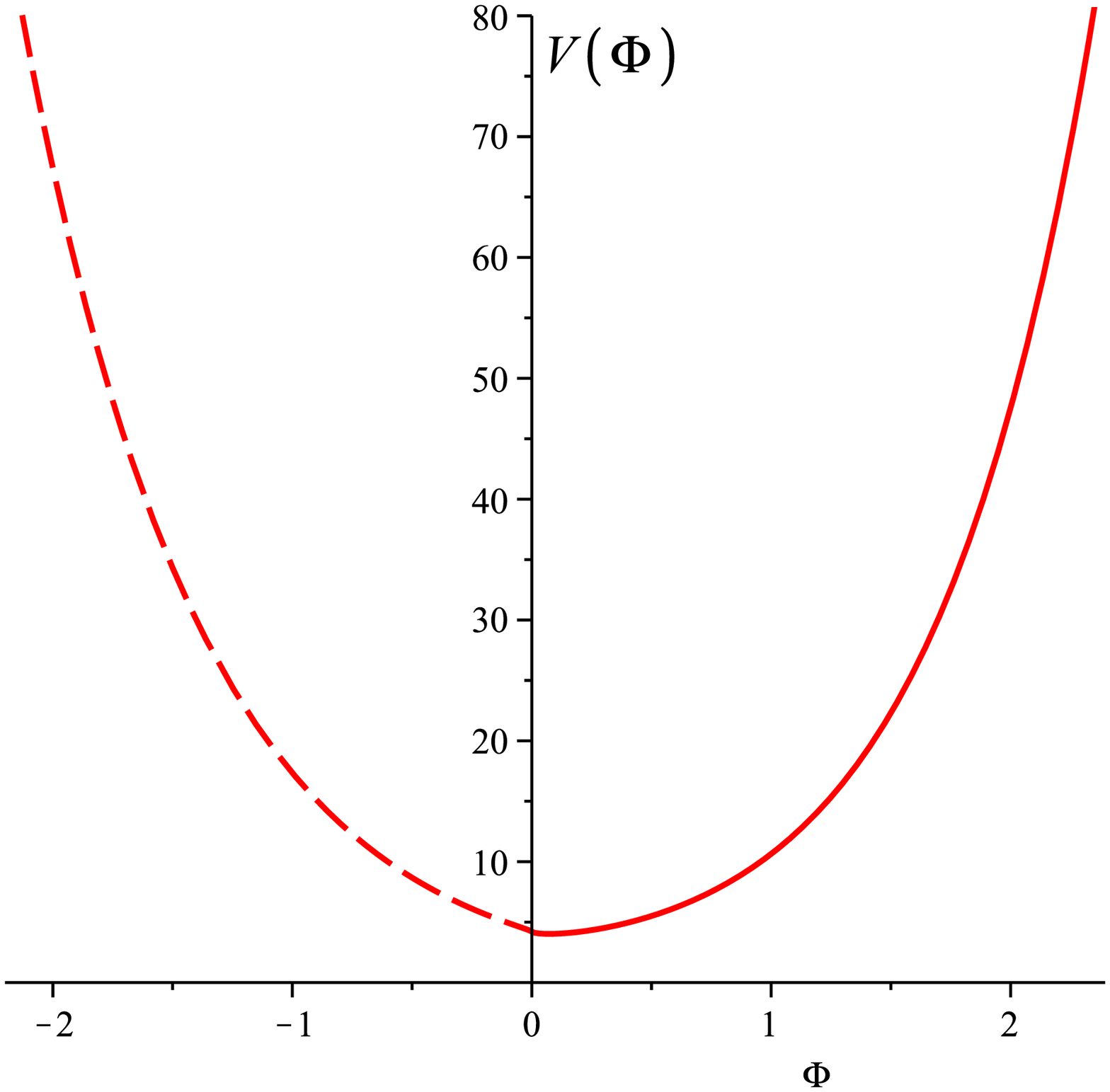}\,B.
\qquad\includegraphics[height=4cm]{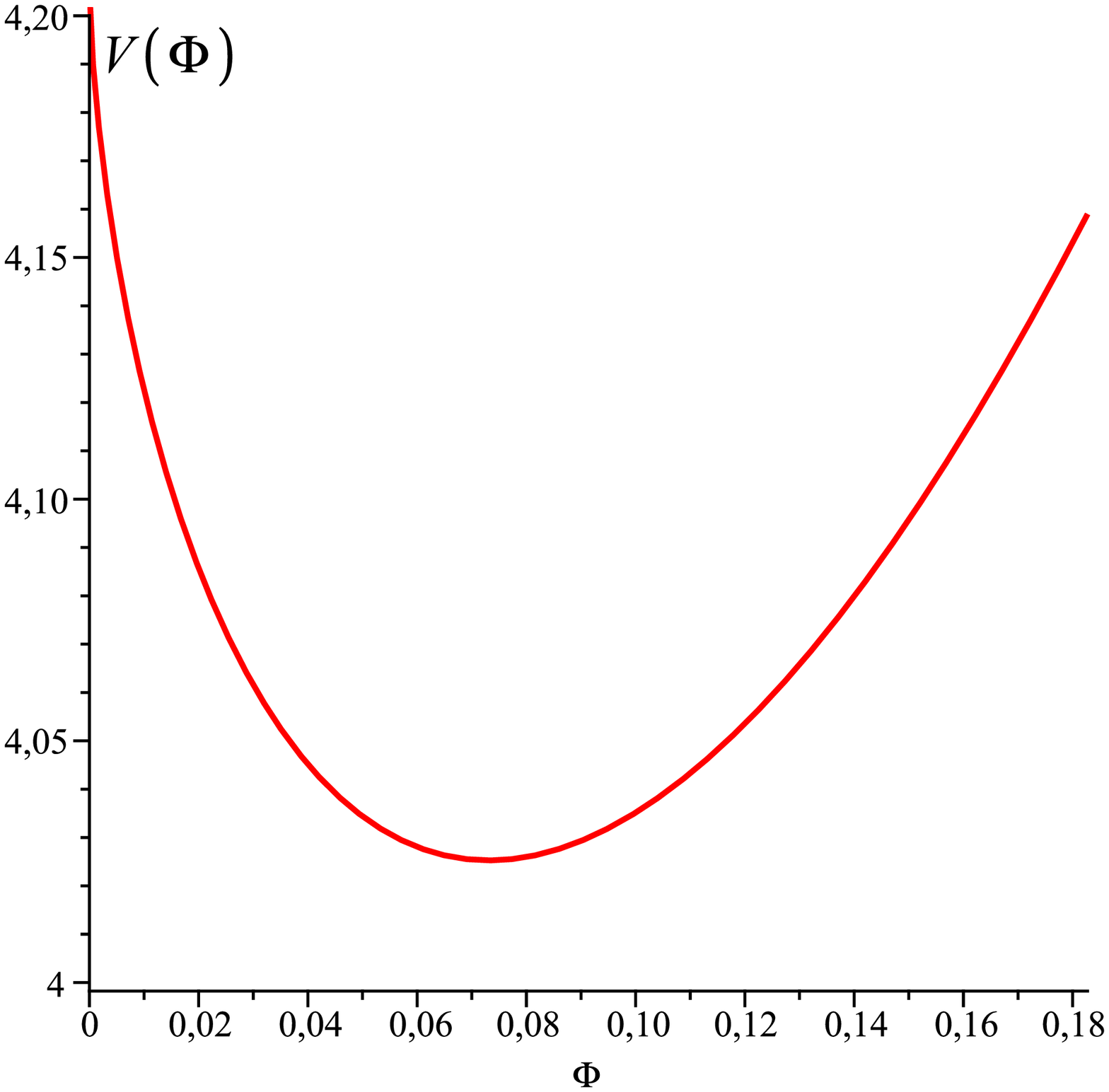}\,C.
\caption{The plots corresponding to $a={1}/{2}$, $L=4.4$ fm, R=1 fm  and a sign plus in \eqref{com}.  A. The phantom  $\Phi_p$ (dashed line) and dilaton $\Phi_s$ (solid line) fields as
functions of $z$. B.  The dependence of the potential $V$ on the dilaton  and  the phantom fields. C. The same dependence of the potential $V$ on the dilaton field  as in B for small $\Phi$.}
\label{odp}
\end{figure}

 The  potential can be  represented parametrically as a functions of the real component $\Phi_s$ for $z>z_0$ and as a function of the imaginary component $\Phi_p$ for $z<z_0$, see Fig. \ref{odp}.

\subsection{Shock wave} \label{ssec2.2}
To deal with a point-like shock wave we add to the action \eqref{ex1} an action of a point-like source moving along a trajectory $x^\mu=x_*^\mu(\eta)$,
\begin{equation*}
 S_{\mathrm{st}}=\int\biggl[\frac{1}{2e}g_{\mu\nu}\frac{dx_*^\mu}{d\eta}\frac{dx_*^\nu}{d\eta}-\frac{e}{2}m^2\biggr]d\eta,
\end{equation*}
$m$ is the particle mass,
 $\eta$ is an arbitrary world-line parameter,we assume that the particle mass is zero, which allows treating only with light-like geodesics.
 $e_\mu^a$ is the frame associated with the metric,
 $g_{\mu\nu}=e^a_{\mu}e_{\nu\,a}$, and $e$ is the square root of its determinant $e=\sqrt{-g}$.
We assume that the metric has the shock wave form \cite{18}--\cite{22}
\begin{equation}
\label{eq3}
 ds^2=b^2(z)\bigl(dz^2+dx^i\,dx^i-dx^{+}\,dx^{-}+\phi(z,x^1,x^2)\delta(x^{+})(dx^{+})^2\bigr),\qquad i=1,2.
\end{equation}
The shock-wave metric solves the Einstein equation
 \begin{equation}
\label{eq2}
 \biggl(R_{\mu\nu}-\frac{g_{\mu\nu}}{2}R\biggr)-\frac{g_{\mu\nu}}{2}\biggl(-\frac{4}{3}(\partial\Phi)^2+V(\Phi)\biggr)-
 \frac{4}{3}\partial_\mu\Phi\,\partial_\nu\Phi-g_{\mu\nu}\frac{d(d-1)}{2L^2}=8\pi G_5 J_{\mu\nu},
\end{equation}
 where $(\partial\Phi)^2=g^{\mu\nu}\partial_{\mu}\Phi\partial_{\nu}\Phi,$ and the current $J_{\mu\nu}$ is given
 by \cite{5}
  \begin{displaymath}
 J_{\mu\nu}=\frac{1}{\sqrt{-g}}\int e p_{\mu}p_{\nu}\delta(x^{\mu}-x_{*}^{\mu})d\eta,
 \end{displaymath}
 here $ p_{\mu}$ is the conjugate momentum
 \begin{displaymath}
 p_{\mu}=e^{-1}g_{\mu\nu}\frac{dx^{\nu}}{d\eta}.
 \end{displaymath}
The current  in light-like coordinates $(x^+,x^-,x^i,z), \,i=1,2$ is written
 by
\begin{equation*}
 J_{++}=\frac{E}{b^3(z)}\,\delta(x^1)\delta(x^2)\delta(z-z_*)\delta(x^{+}).
\end{equation*}
As the metric \eqref{eq3} has the addition of the shock wave profile in comparison with the metric \eqref{ex2}, so we have the  additional  Einstein equation  describing the shock wave profile $\phi(z,x_{\bot})$
\begin{equation}
\label{eq7}
 \biggl(\partial^2_{x^1}+\partial^2_{x^2}+\partial^2_z+\frac{3b^\prime}{b}\partial_z\biggr)\phi(z,x_{\bot})=
 -16\pi G_5\frac{E}{b^3}\,\delta(x^1)\delta(x^2)\delta(z-z_*).
\end{equation}
 Hence, it is  clear  that the dilaton field does not explicitly affect  on the shock wave profile
   resulting from the source.

\subsection{Domain-wall}\label{ssec2.3}
Here,  we consider the Einstein equations for the shock wave resulting from  mass uniformly
distributed over the domain-wall.
The shock wave motion generated by a point mass corresponds to equation \eqref{eq7}.
 To obtain the Einstein equations for the shock waves generated by a domain-wall, we
consider the mass of a point-like source averaged over the domain-wall. Such an averaging method
was proposed in \cite{9}, and we considered it in \cite{ABP}.
To derive the equations of the domain-wall, we use the expression for the induced metric over the wall surface:
\begin{equation}
\label{eq10}
 h_{\alpha\beta}=\frac{\partial x^{\mu}}{\partial {\sigma^{\alpha}}}\frac{\partial x^{\nu}}{\partial {\sigma^{\beta}}}g_{\mu\nu}=b^2\delta_{\alpha\beta}.
\end{equation}
We integrate  \eqref{eq7} over $x_{\bot}=(x_1,x_2)$. According to \eqref{eq10}, one has
\begin{equation*}
 \int \sqrt{h}\,dx_{\bot}=\int b^2\,dx_{\bot},
\end{equation*}
and hence
\begin{equation*}
 \int b^2\biggl( \partial^2_z+ \partial^2_{x^1}+\partial^2_{x^2}+\frac{3b^\prime}{b}\partial_z\biggr)\phi(z,x_{\bot})\,dx_{\bot}=
 -16\pi G_5 b^2\frac{E}{b^3}\delta(z-z_*).
\end{equation*}
Assuming that the derivatives of  $\phi(z,x_{\bot})$ with respect to the transversal variables
 $x_{\bot}$ to be  decrease at  $\pm\infty$, we obtain
the equation of motion for the membrane wall:
\begin{equation*}
 \biggl( \partial^2_z+\frac{3b^\prime}{b}\partial_z\biggr)\phi^{\mathrm W}(z)=-16\pi G_5\frac{E}{b^3}\delta(z-z_*),
\end{equation*}
where
\begin{equation}
\label{eq11}
 \phi^{\mathrm W}(z)=\int\phi(z,x_{\bot})\,dx_{\bot}.
\end{equation}

We can assume the size of the moving domain is finite and  average the mass over the finite surface in \eqref{eq7}.
We consider a wave profile which is provided by the mass uniformly  distributed over the surface perpendicular to the direction of motion.
 Therefore, wave profile \eqref{eq11}
depends on the coordinate  along which the motion occurs, and the equation of the wave profile becomes
 \begin{equation*}
 \biggl(\partial^2_z+\frac{3b^\prime}{b}\partial_z \biggr)\phi^{\omega}(z)=-16\pi G_5 \frac{E}{b^3}\delta(x_{\bot})\delta(z-z_*).
 \end{equation*}
The assumption about the domain is a disk of radius $L$ allows to transform the equation into the form
 \begin{equation*}
 \biggl(\partial^2_z+\frac{3b^\prime}{b}\partial_z\biggr)\phi^{\omega}(z)=
 -16\pi G_5\frac{E^*}{b^3}\delta(z-z_*),\quad\text{���}\; E^*=\frac{E}{L^2}.
\end{equation*}
This shows that the cases of the mass distribution over finite and infinite surfaces are equivalent, i.\,e.\ the profiles are differed by a constant factor
corresponding to the size of the finite object $\phi^{\omega}(z)=\phi^{\mathrm W}(z)/L^2$.

\subsection{Condition of the trapped surface  formation}\label{ssec2.4}
In the  case of $b=L/z$, the conditions on the boundary points $z_a$, $z_b$ of the trapped surface were obtained in  \cite{9,ABP}:
\begin{equation}
\label{eq12}
 (\partial_z\phi^{\omega})\big|_{z=z_a}=2,\qquad (\partial_z\phi^{\omega})\big|_{z=z_b}=-2,
\end{equation}
where  $z_a<z_*<z_b$ is assumed. Obviously,  expressions \eqref{eq12} lead to the condition \footnote{We note that  the condition $(\partial_z\phi^{\omega})^2\big|_{\text{\tiny\rm TS}}=8$ was used in \cite{12}. The difference in the boundary
conditions is associated with the choice of the shock wave metric in \cite{12} in the form
\begin{equation*}
 ds^2=b^2\{dz^2+dx^idx^i-2dx^{+}dx^{-}+\phi(z,x^1,x^2)(dx^{+})^2\}.
\end{equation*}
}
$(\partial_z\phi^{\omega})^2\big|_{\text{\tiny\rm TS}}=4$. Due to $\phi^{W}(z)=L^2\phi^{\omega}(z)$, the condition on the boundary of the trapped surface
formed by the collision of two infinity domain-walls is written as
\begin{equation*}
(\partial_z\phi^{\mathrm W})^2\big|_{\text{\tiny\rm TS}}=4L^4.
\end{equation*}
It is therefore obvious that the values of boundary points of the trapped surface are independent of
 whether the mass is distributed over a finite or infinite
surface.

\section{Power-law $b$-factor}\label{sec3}
We note that the power-law factor of the form $b=(L/(z-z_0))^a$ was used in \cite{12}.
 If $a=1$ and $b=L/(z-z_0)$, then the Einstein equation becomes
\begin{equation}
\label{eq13}
 \biggl(\partial^2_z-\frac{3}{z-z_0}\partial_z\biggr)\phi^{\omega}(z)=-16\pi G_5E^*\frac{(z-z_0)^3}{L^3}\delta(z-z_*).
\end{equation}
Because  $z_0$ is a constant, replacing  $z$ with $z-z_0$  reduces Eq. \eqref{eq13} to the
equation  considered in  \cite{9}. For definiteness in what follows,  we assume that $z_0=0$ in the power-law $b$-factors.

\begin{figure}[h] \centering
\includegraphics[height=7cm]{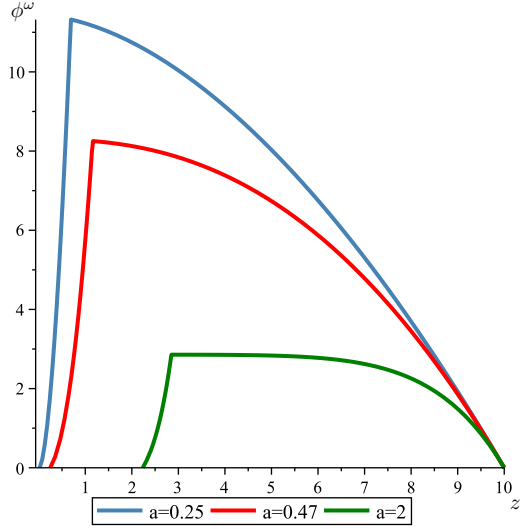}
\caption{The profiles $\phi^{\omega}(z)$ corresponding to $b=\left(\displaystyle\frac{L}{z}\right)^a$ at  $a=0.25$, $a=0.47$, $a=2$, $z_b=10$ fm, $E=0.2$ GeV. The unit of length in Figs. \ref{PowerCase}--\ref{fig4}, \ref{lolo}--\ref{Z_A,Sb} is 1 fm.}\label{PowerCase}
\end{figure}

\begin{figure}[h] \centering
\includegraphics[height=7cm]{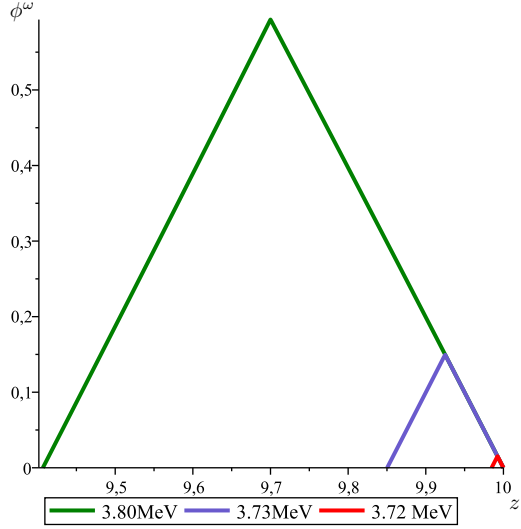}
\caption{The profiles $\phi^{\omega}(z)$ corresponding to $b=\left(\displaystyle\frac{L}{z}\right)^a$ at  $a=0.25$, $z_b=10$ fm, $E=3.72$ MeV, $E=3.73$ MeV, $E=3.80$ MeV.}\label{poweresc}
\end{figure}

The equation of the domain-wall profile in the space with the power-law factor  $b=(L/z)^a$ is written as
\begin{equation}
\label{eq14}
 \biggl(\partial^2_z-\frac{3a}{z}\partial_z\biggr)\phi^{\omega}(z)=-16\pi G_5\biggl(\frac{z}{L}\biggr)^{3a}E^*\delta(z-z_*).
\end{equation}
We consider this equation separately before and after the collision.
The boundary points of trapped surface of the black hole are denoted by
$z_a$ and $z_b$, $z_a<z_*<z_b$. The solution of \eqref{eq14}
is written as
\begin{equation}
\label{eq15}
 \phi^{\omega}(z)=\phi^{\omega}_a\Theta(z_*-z)+\phi^{\omega}_b\Theta(z-z_*),
\end{equation}
where
\begin{align*}
&\phi^{\omega}_a(z)=C_0z_az_b\biggl(\!\biggl(\frac{z_*}{z_b}\biggr)^{3a+1}-1\biggr)\biggl(\!\biggl(\frac{z}{z_a}\biggr)^{3a+1}-1\biggr),
\\
&\phi^{\omega}_b(z)=C_0z_az_b\biggl(\!\biggl(\frac{z_*}{z_a}\biggr)^{3a+1}-1\biggr)\biggl(\!\biggl(\frac{z}{z_b}\biggr)^{3a+1}-1\biggr),
\\
&C_0=-\frac{16\pi G_5Ez_a^{3a}z_b^{3a}}{(1+3a)L^{3a+2}(z_b^{3a+1}-z_a^{3a+1})}.
\end{align*}
The given profile is presented  in Figs. \ref{PowerCase}, \ref{poweresc}.

For profile \eqref{eq15}, the formation conditions for the trapped surface at the boundary points  $z=z_a$ and $z=z_b$  are
\begin{equation}
\label{eq16}
\begin{aligned}
&\frac{8\pi G_5Ez_a^{3a}(1-z_b^{3a+1}/{z_*}^{3a+1})}{L^{3a+2}(z_b^{3a+1}/{z_*}^{3a+1}-z_a^{3a+1}/{z_*}^{3a+1})}=-1,
\\
&\frac{8\pi G_5Ez_b^{3a}(1-z_a^{3a+1}/{z_*}^{3a+1})}{L^{3a+2}(z_b^{3a+1}/{z_*}^{3a+1}-z_a^{3a+1}/{z_*}^{3a+1})}=1.
\end{aligned}
\end{equation}

\begin{figure}[t]
\centering
\includegraphics[height=6cm]{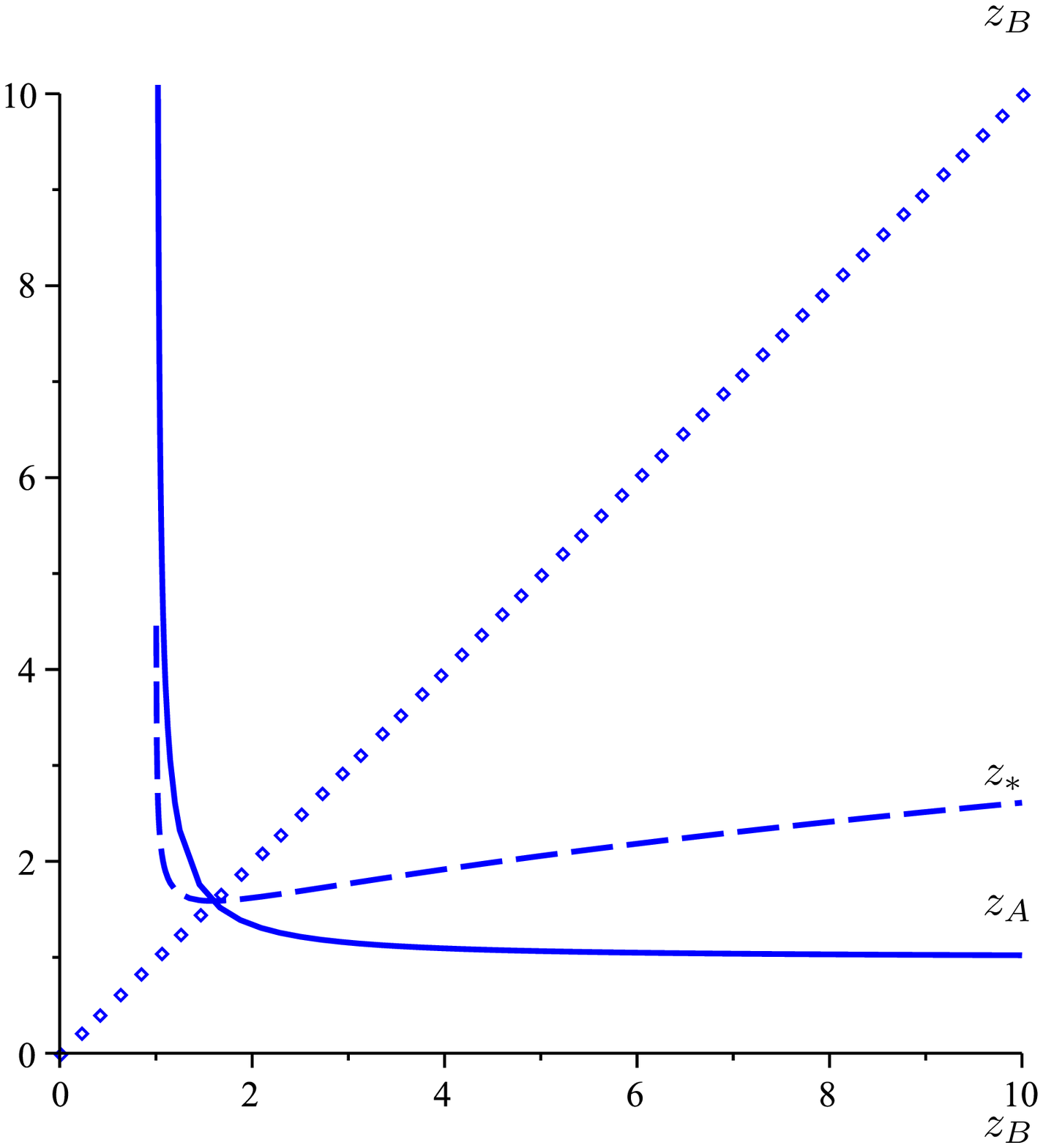}
\caption{The solution to the system of equations  \eqref{eq16} for the given $z_b$.
\label{fig1}}
\end{figure}

The collision point $z_{\ast}$ can not be fixed but found from the system regarded as a system of equations for    $z_*$, $z_a$ with a given $z_b$:
\begin{equation}
\label{eq17}
 z_a=\biggl(\frac{z_b^{3a}}{-1+z_b^{3a}C}\biggr)^{1/3a},\qquad
 z_*=\biggl(\frac{z_a^{3a}z_b^{3a}(z_b+z_a)}{z_a^{3a}+z_b^{3a}}\biggr)^{1/(3a+1)}\,,
\end{equation}
where $C=8\pi G_5E/L^{3a+2}$.  The solution of this system is shown in Fig. \ref{fig1}.

  As a consequence of the condition $z_{a}<z_{b}$ and (\ref{eq17}), we obtain the range of the energies:
 $E>L^{3a+2}/{4\pi G_{5}z_{b}^{3a}}$ at which the  black hole profile is created. It is clear that the  black hole profile is not created at the energy $E=L^{3a+2}/{4\pi G_{5}z_{b}^{3a}}$, and for the parameters $G_{5} = L^{3}/1.9$, $L = 4.4$ fm and $z_{b}=0.5$ fm it equals to
  $E=3.71 $ MeV. (see. Fig. \ref{poweresc}).

For $z_b^{3a}C\gg 1$, we consider $z_a\ll z_*\ll z_b$. Based on  \eqref{eq17}, we have the approximation:
\begin{equation*}
 z_a\sim\biggl(\frac{1}{C}\biggr)^{1/3a},\qquad z_*\sim\biggl(\frac{z_b}{C}\biggr)^{1/(3a+1)}.
\end{equation*}
The trapped surface area is calculated as
\begin{equation*}
 S_{\text{\text{trap}}}=\frac{1}{2G_5}\int_C\sqrt{\det|g^{}_{\text{AdS}_3}|}\,dz\,d^2x_\perp,
\end{equation*}
where $\det|g^{}_{\text{AdS}_3}|$ is the metric determinant of free dimensional space  $\mathrm{AdS}_3$. In the follows,
 we calculate the relative area $s$ of the trapped surface defined by
\begin{equation*}
 s=\frac{S_{\text{trap}}}{\int d^2x_{\perp}}=\frac{1}{2G_5}\int^{z_b}_{z_a}b^{3}\,dz.
\end{equation*}
In the considered  case where  $b(z)=(L/z)^a$, the formula for the relative area of trapped surface becomes
\begin{equation*}
 s=\frac{1}{2G_5(3a-1)}\biggl(z_a \biggl(\frac{L}{z_a}\biggr)^{3a}-z_b\biggl(\frac {L}{z_b}\biggr)^{3a}\biggr),
\end{equation*}
and $s$ determines the relative entropy. With the assumption $3a>1$ and the used approximation,  it is clear from this expression that the relative area of the trapped surface tends to its maximum value at infinite $z_b$:
\begin{equation}
\label{eq18}
 s|_{z_b\to\infty}=\frac{L^{3a}}{2G_5(3a-1)}z_a^{1-3a}=\frac{L}{2G_5}\biggl(\frac{8\pi G_5}{L^2}\biggr)^{(3a-1)/3a}E^{(3a-1)/3a}.
\end{equation}
We thus find  that for $a>1/3$, the  entropy $S$ increases as $E^{(3a-1)/3a}$.

 We substitute parameters and variables with the dimension of length in formula \eqref{eq18} using the relation $1\,\text{GeV}\approx5\,\text{fm}^{-1}$ and
   choose the parameters $G_5$ and $L$  based on  phenomenological reasons \cite{5}: $G_5=L^3/1.9$ and $L=4.4$\,fm. We here assume that we consider  collisions of lead ions.
The multiplicity of particles produced in heavy-ion  collisions (Pb-Pb and Au-Au collisions)
depends on energy  as  $s^{0.15}_{{}_{NN}}$ according to the experimental data ~\cite{17}
 in the  range from  $10$ to $10^3$\, GeV. Therefore, $a\approx 0.47$. For $a = 0.47$, we have
\begin{equation*}
 \frac{L}{2G_5}\biggl(\frac{8\pi G_5}{L^2}\biggr)^{(3a-1)/3a}\approx 0.16\,\text{fm}^{-1.85}.
\end{equation*}

\section{Factor of the form $b={\mathrm e}^{-z/R}$}\label{sec4}
\begin{figure}[t]
\centering
\includegraphics[height=5cm]{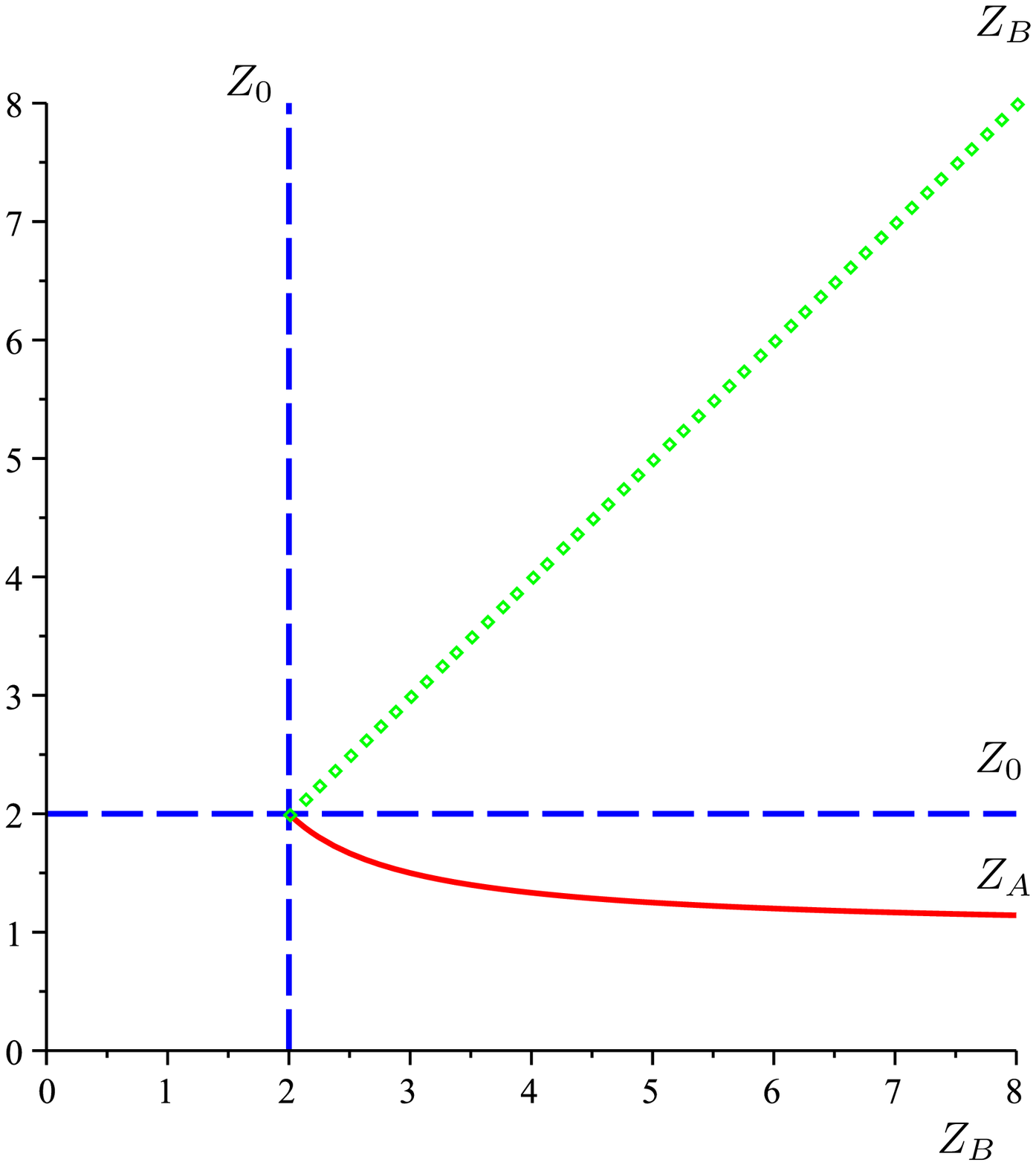}\,\,\,A.\,\,\,\,\,\,\,\,\,\,\,\,\,\,\,\,\,\,\,\,\,\,\,\,
\includegraphics[height=5cm]{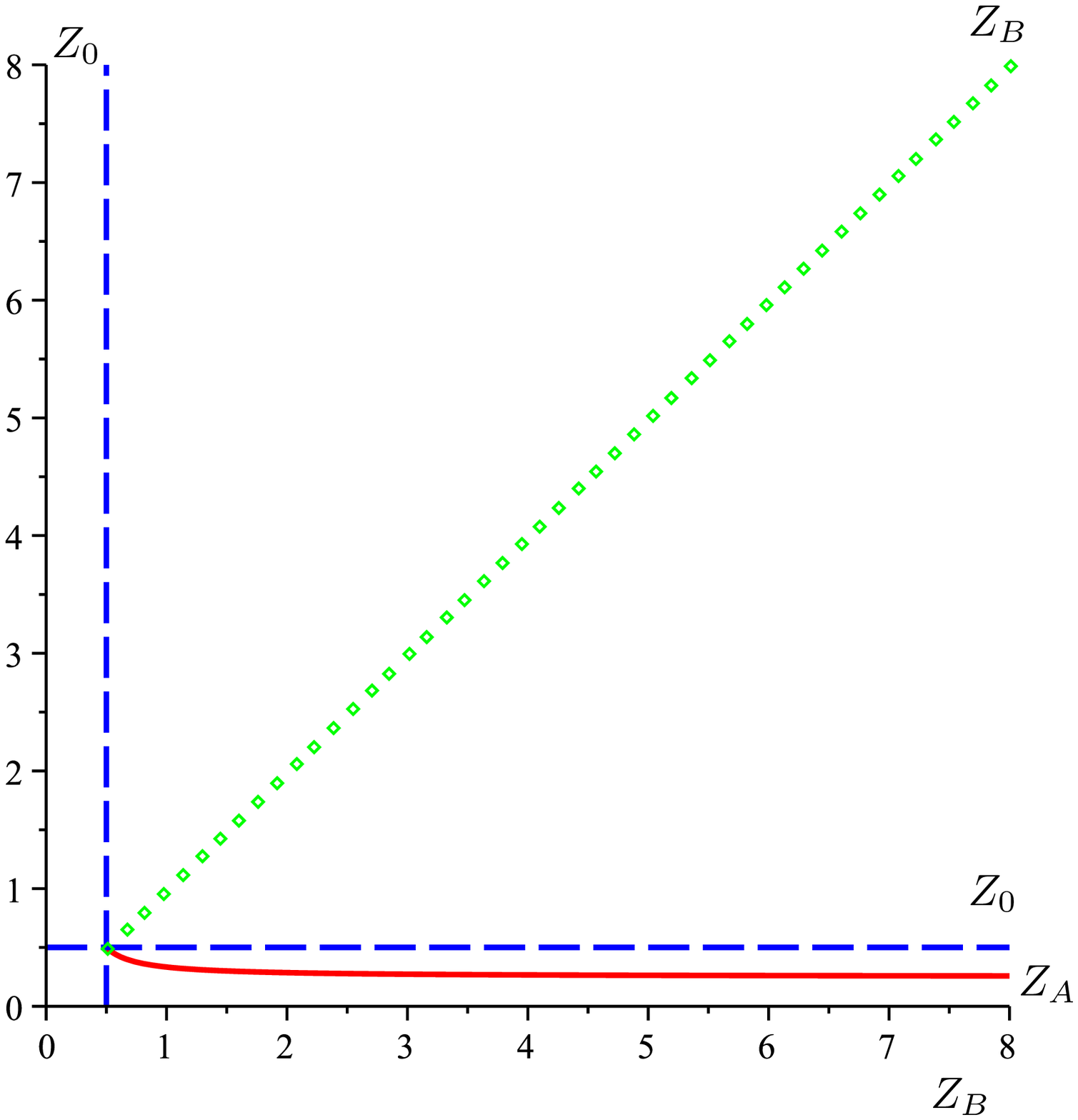} B.
\caption{Solution \eqref{eq20} at $Z_*=4$ and ~(b) $Z_*=2$: the vertical dashed line separates the region where $Z_b>Z_*$.}\label{fig2}
\end{figure}

\begin{figure}[h] \centering
\includegraphics[height=7cm]{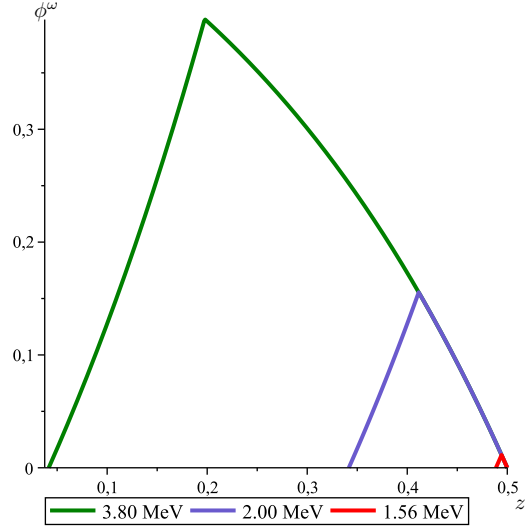}
\caption{ The profiles $\phi^w(z)$ corresponding to $b={\mathrm e}^{-z/R}$ at $z_b=0.5$ fm, $R=1$ fm, $E=1.56$ MeV, $E=2.00$ MeV, $E=3.80$ MeV.}\label{ExpCase}
\end{figure}

The equation of the domain-wall wave profile in the space with the exponential $b$-factor $b={\mathrm e}^{-z/R}$ is written as
\begin{equation*}
 \biggl(\partial^2_z-\frac{3}{R}\partial_z\biggr)\phi^{\omega}(z)=-16\pi G_5E^*{\mathrm e}^{3z/R}\delta(z-z_*),
\end{equation*}
and we construct the solution in the form
\begin{equation*}
 \phi^{\omega}(z)=\phi_a(z)\Theta(z_*-z)+\phi_b(z)\Theta(z-z_*),
\end{equation*}
where
\begin{alignat*}{3}
&\phi_a=C_a\frac{R}{3}{\mathrm e}^{3z/R}+\widetilde C_a,&\qquad &\phi_b=C_b\frac{R}{3}{\mathrm e}^{3z/R}+\widetilde C_b,
\\
&C_a=-\frac{16\pi G_5E^*\bigl({\mathrm e}^{3z_*/R}-{\mathrm e}^{3z_b/R}\bigr)}{{\mathrm e}^{3z_b/R}-{\mathrm e}^{3z_a/R}},&\qquad
&C_b=-\frac{16\pi G_5E^*\bigl({\mathrm e}^{3z_*/R}-{\mathrm e}^{3z_a/R}\bigr)}{{\mathrm e}^{3z_b/R}-{\mathrm e}^{3z_a/R}},
\\
&\widetilde C_a=-C_a\frac{R}{3}{\mathrm e}^{3z_a/R},&\qquad &\widetilde C_b=-C_b\frac{R}{3}{\mathrm e}^{3z_b/R}.
\end{alignat*}
The conditions at the boundaries of trapped surface are
\begin{equation}
\label{eq19}
\begin{aligned}
 \frac{8\pi G_5E}{L^2}\frac{\bigl({\mathrm e}^{3z_*/R}-{\mathrm e}^{3z_b/R}\bigr){\mathrm e}^{3z_a/R}}{{\mathrm e}^{3z_b/R}-{\mathrm e}^{3z_a/R}}&=-1,
\\
 \frac{8\pi G_5E}{L^2}\frac{\bigl({\mathrm e}^{3z_*/R}-{\mathrm e}^{3z_a/R}\bigr){\mathrm e}^{3z_b/R}}{{\mathrm e}^{3z_b/R}-{\mathrm e}^{3z_a/R}}&=1.
\end{aligned}
\end{equation}
We analyze these conditions. We set
\begin{equation}\label{notations}
 Z_*={\mathrm e}^{3z_{*}/R},\qquad Z_a={\mathrm e}^{3z_a/R},\qquad Z_b={\mathrm e}^{3z_b/R}.
 \end{equation}
 and substitute these values in conditions \eqref{eq19}. We obtain the equations
\begin{equation*}
\frac{8\pi G_5E}{L^2}\frac{(Z_*-Z_b)Z_a}{Z_b-Z_a}=-1,\qquad \frac{8\pi G_5E}{L^2}\frac{(Z_*-Z_a)Z_b}{Z_b-Z_a}=1.
\end{equation*}
As in the previous case, we consider the equations for $Z_a$ and $Z_*$ with a fixed $Z_b$. \\
This system has the trivial solution $Z_a=Z_b=Z_*$ and the solution
\begin{equation}
\label{eq20}
 Z_a=\frac{L^2}{8\pi G_5E}\frac{Z_b}{Z_b-L^2/8\pi G_5E}\,,\qquad Z_*=\frac{L^2}{4\pi G_5E}.
\end{equation}
This solution is shown in Fig. ~\ref{fig2}  at two values of the parameter $Z_*$:
\begin{equation*}
 \frac{Z_{*}}{4}=1,\qquad \frac{Z_{*}}{4}=\frac{1}{2}.
\end{equation*}
The relative area of trapped surface is given by
\begin{equation}
\label{eq21}
 s=\frac{R}{6G_5}\biggl(\frac{1}{{\mathrm e}^{3z_a/R}}-\frac{1}{{\mathrm e}^{3z_b/R}}\biggr)=
 \frac{R}{6G_5}\biggl(\frac{1}{Z_a}-\frac{1}{Z_b}\biggr).
\end{equation}
The maximum entropy is attained  for $Z_b\gg 1$. In this approximation,
\begin{equation}
\label{eq22}
 Z_a\sim \frac{L^2}{8\pi G_5E},\qquad s\sim\frac{4}{3}\frac{\pi E R }{L^2}.
\end{equation}
According to (\ref{eq20}) and (\ref{notations}) we get the colliding point and the boundary one  $z_a$, namely,
\begin{equation}
  z_{*}=\frac{R}{3}\ln\left(   \frac{L^2}{4\pi E G_5} \right),
\end{equation}
\begin{equation}
z_a=\frac{R}{3}\ln\left( \frac{e^{ {3z_b}/{R}}}{\frac{8\pi  G_5 E}{L^2} e^{ {3z_b}/{R}}-1} \right).
\end{equation}
Based on the condition that $z_a<z_b$,  the lowest energy limit is found, and due to the condition  $z_a>0$,   the highest
energy limit is defined, so
\begin{displaymath}
 \frac{L^{2}}{4\pi G_{5}}e^{-3z_{b}/R} <E<\frac{L^2}{8\pi G_{5}}\left( 1+e^{-3z_{b}/R}\right).
\end{displaymath}
For the parameters $G_{5} = L^{3}/1.9$, $L = 4.4$ fm and $z_{b}=0.5$ fm it takes the values

$$  1.53 \text{MeV}<E< 4.20 \text{MeV}.$$
The corresponding profile $\phi^{\omega}$ is represented in Fig. \ref{ExpCase}.

\section{Mixed  factor of the form $b=(L/z){\mathrm e}^{-z^2/R^2}$}\label{sec5}
The equation of the profile corresponding to the domain-wall motion in the space with  $b$-factor of the form ~ $b=(L/z){\mathrm e}^{-z^2/R^2}$ is written as
\begin{equation*}
 \biggl(\partial^2_z-3\biggl(\frac{1}{z}+\frac{2z}{R^2}\biggr)\partial_z\biggr)\phi^{\omega}=
 -16\pi G_5E^*\biggl(\frac{z}{L}\biggr)^{3}{\mathrm e}^{3z^2/R^2}\delta(z-z_*).
\end{equation*}
We consider the solution of the obtained equation:
\begin{equation}
\label{eq23}
 \phi^{\omega}=\phi^{\omega}_a\Theta(z_*-z)+\phi^{\omega}_b\Theta(z-z_*),
\end{equation}
where
\begin{align*}
&\phi_a^{\omega}=-C_a(R^2-3z_a^2){\mathrm e}^{3z_a^2/R^2}+C_a(R^2-3z^2){\mathrm e}^{3z^2/R^2},
\\
&\phi_b^{\omega}=-C_b(R^2-3z_b^2){\mathrm e}^{3z_b^2/R^2}+C_b(R^2-3z^2){\mathrm e}^{3z^2/R^2},
\end{align*}
and the factors $C_a$ and $C_b$ are given by
\begin{equation}
\label{eq24}
\begin{aligned}
&C_a=C_0\bigl(-(R^2-3z_b^2){\mathrm e}^{3z_b^2/R^2}+(R^2-3z_*^2){\mathrm e}^{3z_*^2/R^2}\bigr),
\\
&C_b=C_0\bigl(-(R^2-3z_a^2){\mathrm e}^{3z_a^2/R^2}+(R^2-3z_*^2){\mathrm e}^{3z_*^2/R^2}\bigr),
\end{aligned}
\end{equation}
where
\begin{equation}
\label{eq25}
 C_0=\frac{8\pi G_5E^*R^2}{9L^3\bigl((R^2-3z_b^2){\mathrm e}^{3z_b^2/R^2}-(R^2-3z_a^2){\mathrm e}^{3z_a^2/R^2}\bigr)}.
\end{equation}
Differentiating each of the two terms in the right-hand sides of solution ~\eqref{eq23}, we get
\begin{equation*}
 \frac{d\phi^{\omega}_a(z)}{dz}=-C_a\frac{18z^3}{R^2}{\mathrm e}^{3z^2/R^2},\qquad
 \frac{d\phi^{\omega}_b(z)}{dz}=-C_b\frac{18z^3}{R^2}{\mathrm e}^{3z^2/R^2}.
\end{equation*}
Hence, the conditions for the trapped surface formation at the boundary points become
\begin{equation}
\label{eq26}
 C_a\frac{9z_a^3}{R^2}{\mathrm e}^{3z_a^2/R^2}=-1,\qquad C_b\frac{9z_b^3}{R^2}{\mathrm e}^{3z_b^2/R^2}=1.
\end{equation}
Considering this system with $C_a$ and $C_b$ defined by formulas  \eqref{eq24} and \eqref{eq25}, we obtain a system of two equations for the three unknowns $z_a$, $z_b$ and  $z_{*}$.  We assume that $z_a$ and $z_{*}$ are unknown and $z_b$ is given.

From system of equations ~\eqref{eq26}, we can  obtain the relations
\begin{align*}
&z_a^3{\mathrm e}^{3z_a^2/R^2}=\frac{L^3z_b^3{\mathrm e}^{3z_b^2/R^2}}{8\pi G_5E^*z_b^3{\mathrm e}^{3z_b^2/R^2}-L^3},
\\*
&(R^2-3z_*^2){\mathrm e}^{3z_*^2/R^2}=
\frac{(z_a^3R^2-3z_a^3z_b^2-3z_b^3z_a^2+z_b^3R^2){\mathrm e}^{3z_a^2/R^2}{\mathrm e}^{3z_b^2/R^2}}{z_a^3{\mathrm e}^{3z_a^2/R^2}+z_b^3{\mathrm e}^{3z_b^2/R^2}}.
\end{align*}

We  consider only energies satisfying  $8\pi G_5E^*z_b^3{\mathrm e}^{3z_b^2/R^2}>L^3$.
 For the further analysis of the trapped surface formation, we must obtain the solution of system ~ \eqref{eq26} ~ in an explicit form. A nontrivial solution of the system (we recall that $z_a$,
$z_{*}$ and ~ $z_b$ are positive) has the form
\begin{equation}
\label{eq27}
\begin{aligned}
&z_a=\frac{R}{\sqrt{2}}\sqrt{W_{\mathrm A}},
\\
&z_*=\frac{R}{\sqrt{3}}\sqrt{1+W
\biggl(\frac{{-(z_a^3R^2-3z_a^3z_b^2-3z_b^3z_a^2+z_b^3R^2)\mathrm e}^{3(z_a^2+z_b^2)/R^2}}
{R^2(z_a^3{\mathrm e}^{3z_a^2/R^2}+z_b^3{\mathrm e}^{3z_b^2/R^2}){\mathrm e}}\biggr)},
\end{aligned}
\end{equation}
where
\begin{equation}
\label{eq28}
 W_{\mathrm A}=W\biggl(2\biggl(\frac{(L^3z_b^3/R^3){\mathrm e}^{3z_b^2/R^2}}{8\pi G_5E^*z_b^3{\mathrm e}^{3z_b^2/R^2}-L^3}\biggr)^{2/3}\biggr),
\end{equation}
and $W(z)$ is Lambert  $W$-function.

\begin{figure}[t!]
\centering
\includegraphics[height=4.7cm]{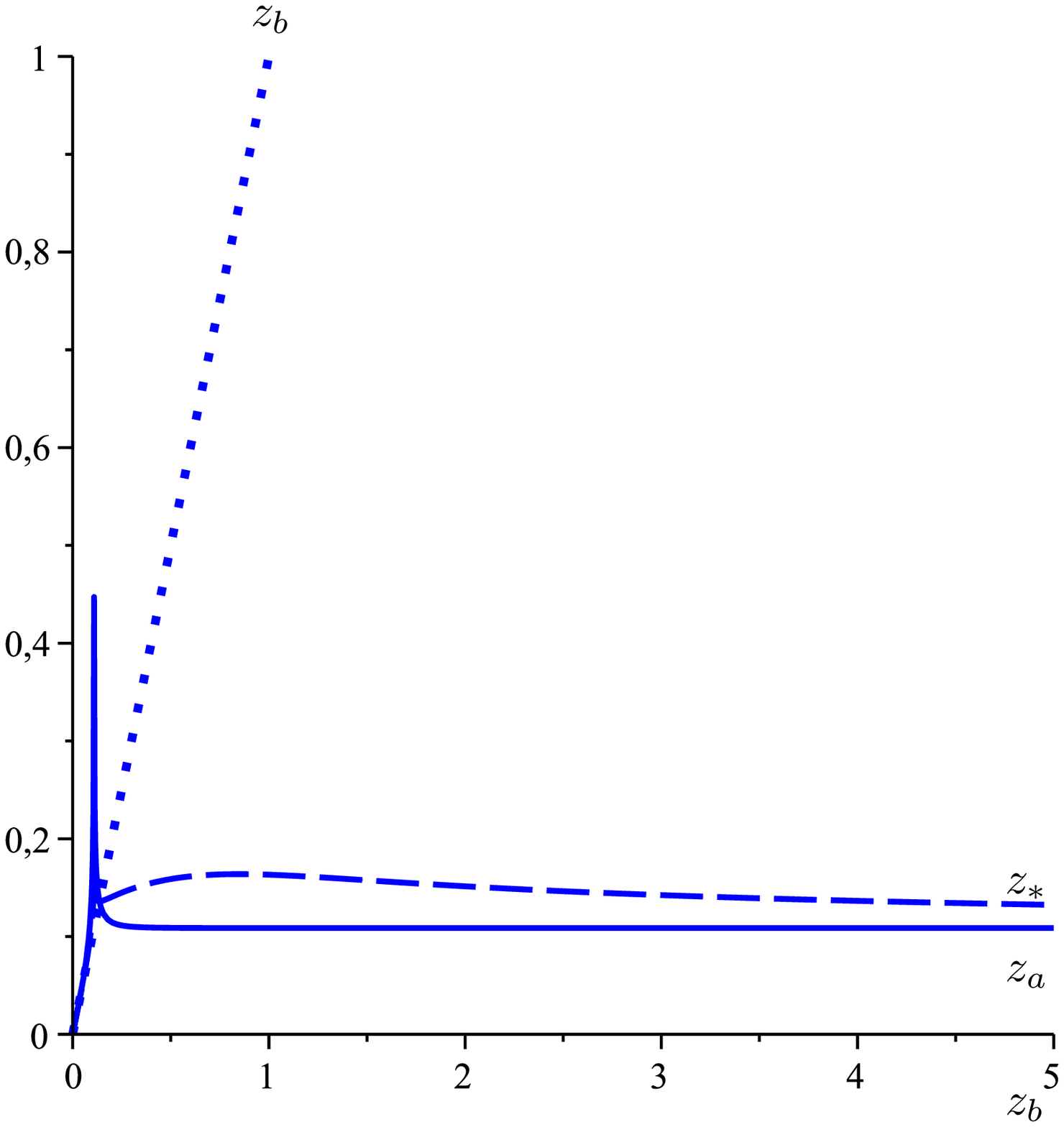}\,A.$\quad$
\includegraphics[height=4.7cm]{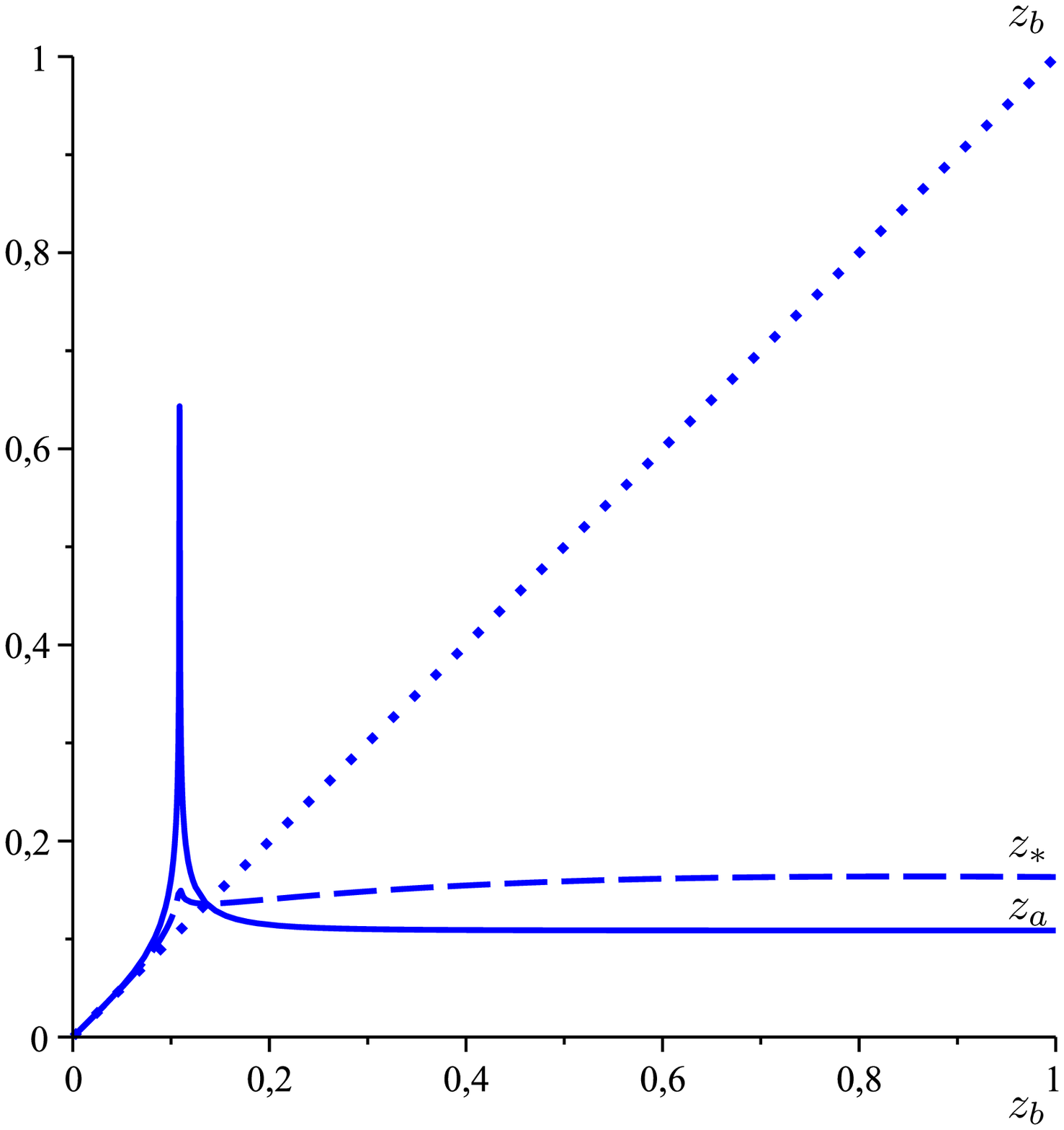}\,B.$\quad$
\includegraphics[height=4.7cm]{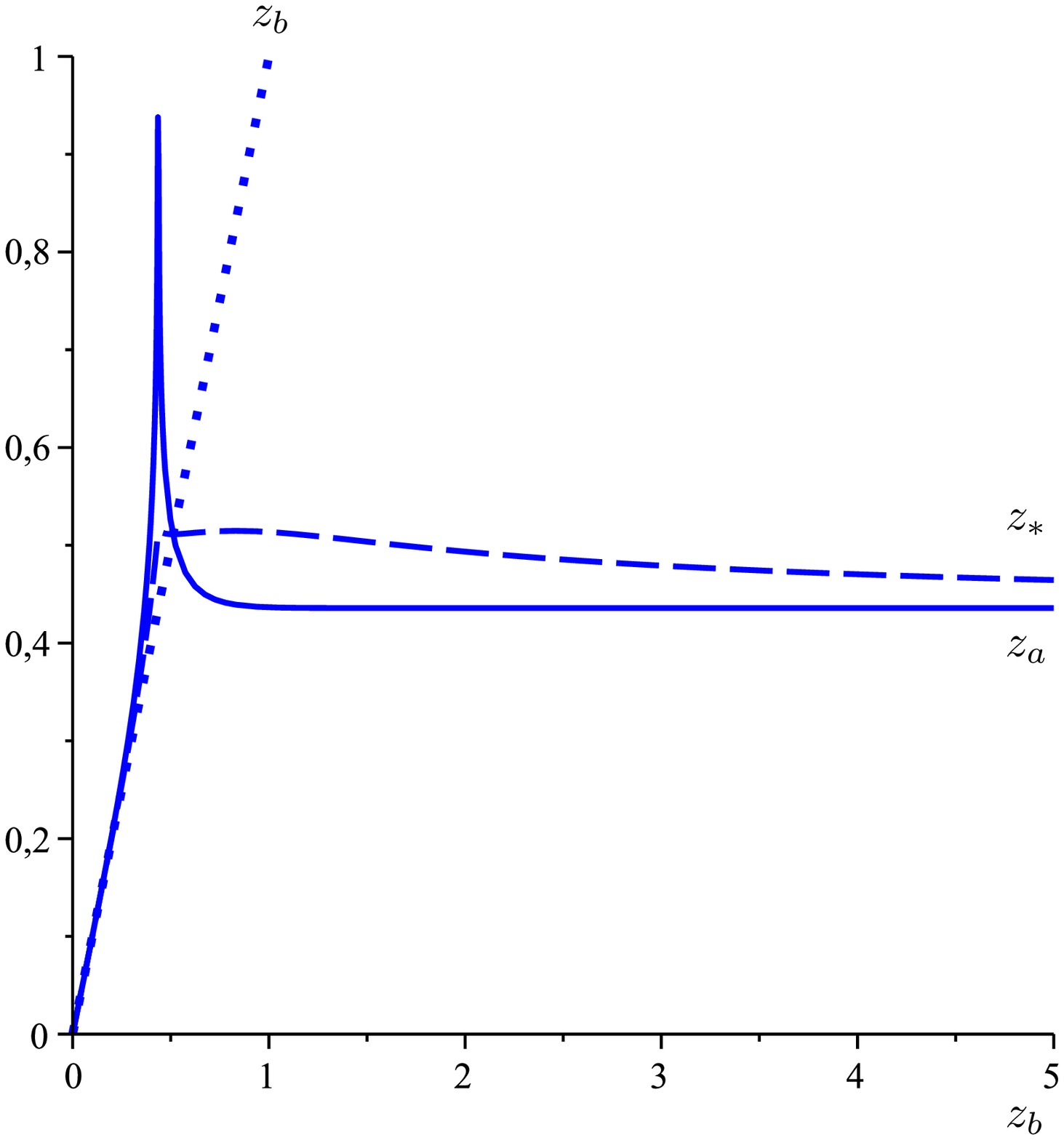}\,C.
\caption{The dependence  $z_a$, $z_{*}$ on $z_b$ at energies $E =220$ \,GeV ~ (Fig. A. at $z_b\leq 5$ fm, Fig. B. at $z_b\leq 1$ fm) and  $E=2$ \,GeV~( Fig. C.).
\label{fig3}}
\end{figure}
\begin{figure}[t] \centering
\includegraphics[height=7cm]{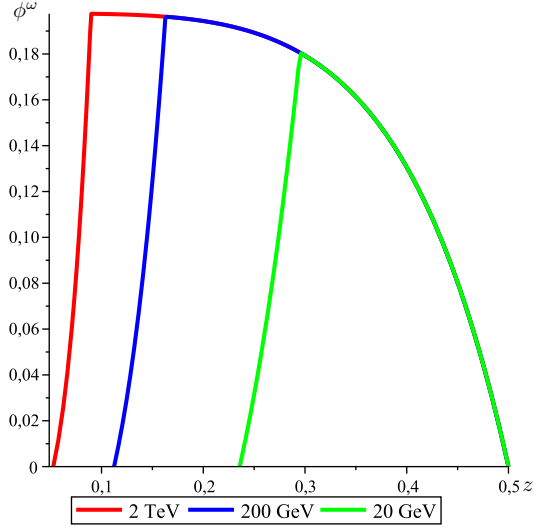}
\caption{The profiles $\phi^{\omega}$ corresponding to  $b(z)=\left(\displaystyle\frac{L}{z}\right)\exp\left(-\frac{z^2}{R^2}\right)$ at $z_b=0.5$ fm, $R=1$ fm, $E=20$ GeV, $E=200$ GeV, $E=2000$ GeV. }\label{Mixed1}
\end{figure}

The dependencies  of  $z_a$ and $z_{*}$ on $z_b$ defined by formulas \eqref {eq27}, \eqref {eq28}, are shown in Figs. \ref{fig3}A. and \ref{fig3}B. at the respective energy values $E=220$\,GeV and $E=2$\, GeV.
We see that $z_a$ tends to its the lowest value  at  infinitely large $z_b$.
In this limit $z_a$ and $z_{*}$ are given by
\begin{equation}
\label{eq29}
 z_a|_{z_b\to\infty}=\frac{R}{\sqrt{2}}\sqrt{W_{\mathrm{AM}}},\qquad
 z_{*}|_{z_b\to\infty}=\frac{R}{\sqrt{3}}\sqrt{1+W\biggl(\frac{(3z_a^2-R^2){\mathrm e}^{3z_a^2/R^2}}{{\mathrm e}R^2}\biggr)},
\end{equation}
where
\begin{equation*}
 W_{\mathrm{AM}}= W\biggl(\frac{L^2}{2(\pi G_5E^*)^{2/3}R^2}\biggr).
\end{equation*}
 In this case, the profile is presented in Fig. (\ref{Mixed1}).

\begin{figure}[t] \centering
\includegraphics[height=7cm]{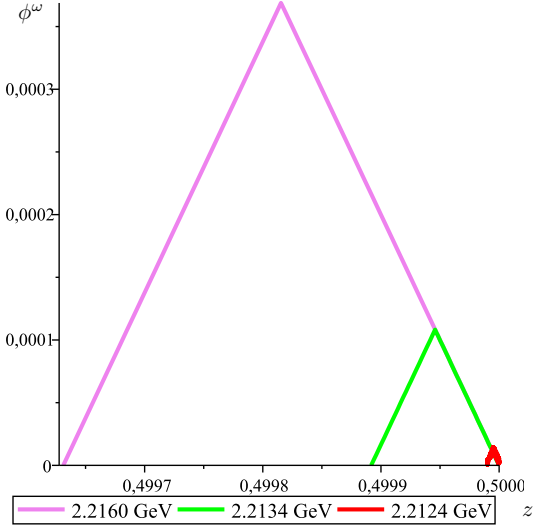}
\caption{The profiles $\phi^{\omega}$ corresponding to  $b(z)=\left(\displaystyle\frac{L}{z}\right)\exp\left(-\displaystyle\frac{z^2}{R^2}\right)$ at $z_b=0.5$ fm, $R=1$ fm, $E=2.2160$ GeV, $E=2.2134$ GeV, $E=2.2160$ GeV. }\label{n=1law}
\end{figure}

 Due to the condition $z_a<z_b$ we found the range of the energies
 at which black hole is created
\begin{displaymath}
E>\frac{L^{5}}{4\pi G_{5}z_{b}^{3}}\exp\left( - \frac{3z_{b}^{2}}{R^{2}} \right).
\end{displaymath}
Black hole is not created already at the energy
\begin{equation}\label{Edis1}
  E= \frac{L^{5}}{4\pi G_{5} zb^{3}}\exp\left( -\frac{3z_{b}^2}{R^{2}}\right) .
\end{equation}
 For the parameters $G_{5} = L^{3}/1.9$, $L = 4.4$ fm and $z_{b}=0.5$ fm it equals to $2.2123$ GeV  (see Fig. \ref{n=1law}).

\begin{figure}[t]
\centering
\includegraphics[height=5cm]{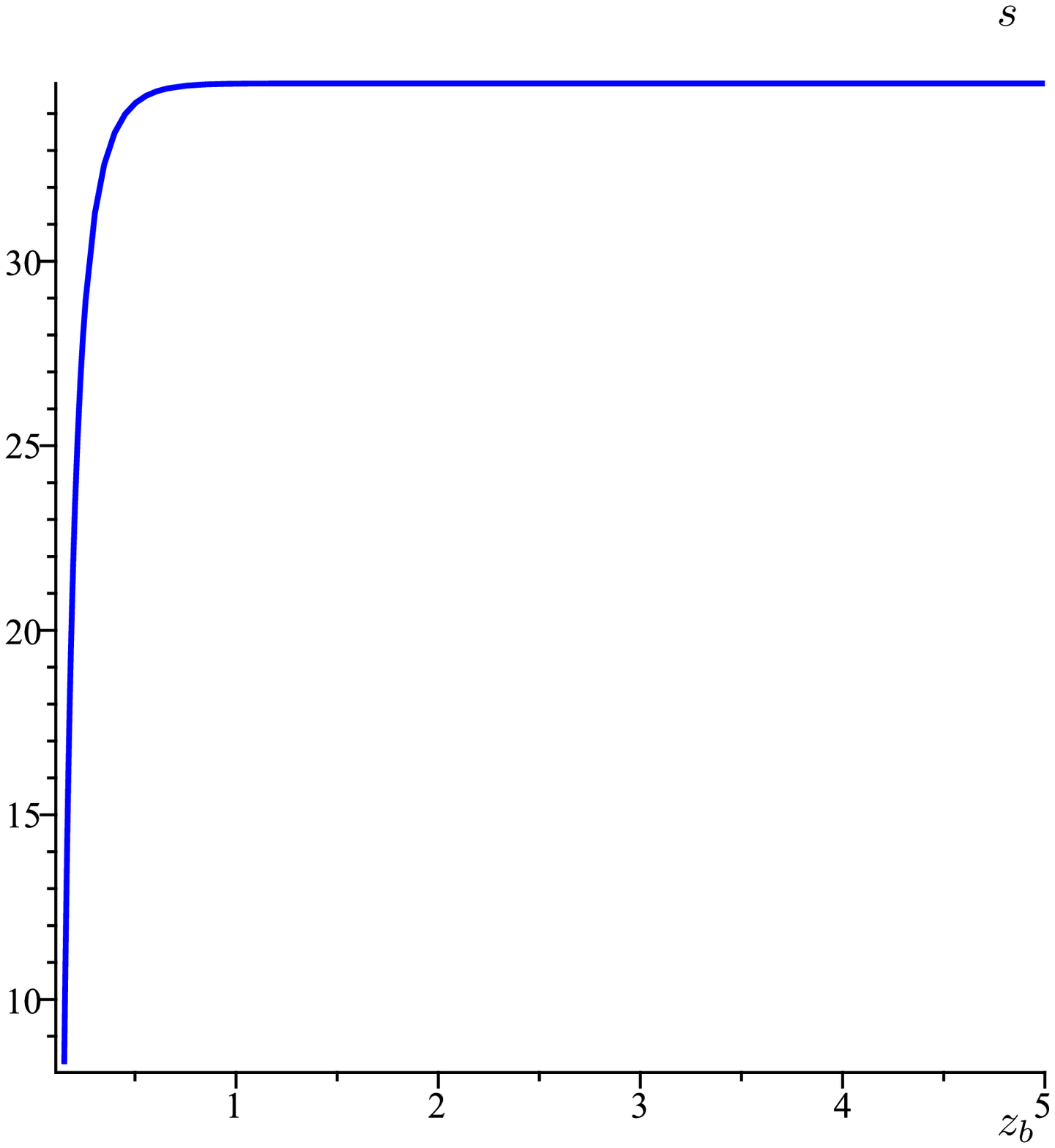}\,A.$\quad$
\includegraphics[height=5cm]{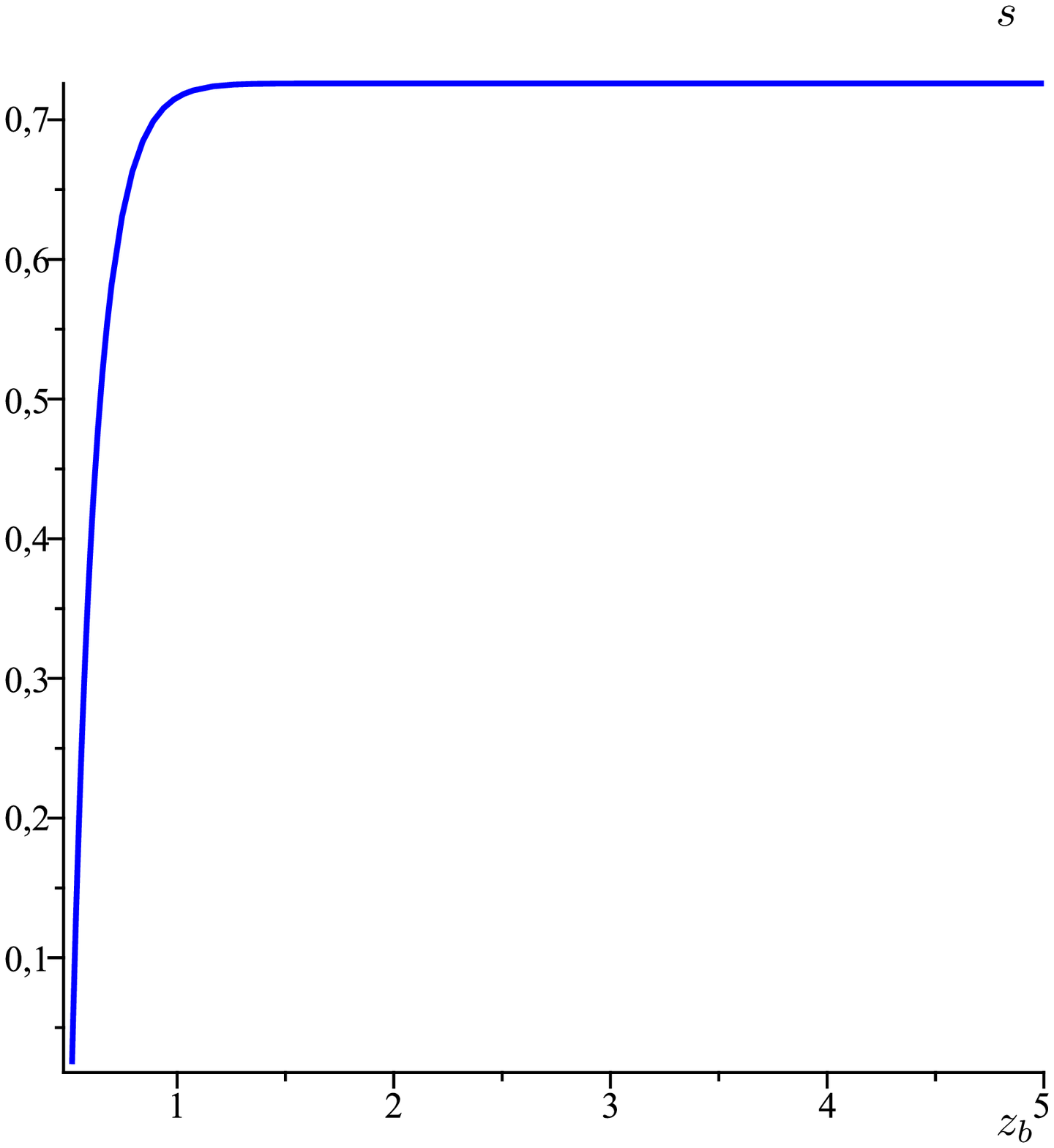}\,B.
\caption{The dependence  $s$ on $ z_b $ at ~ $E=220$\,GeV (A.) and ~ $E=2$\, GeV (B.)\label{fig4}.}
\end{figure}

In general, the relative area of the trapped surface  depends on the energy and on $z_b$ as
\begin{equation*}
 s = \frac{L^3}{2G_5}\left(-\frac{1}{2\exp\left(\displaystyle\frac{3z_b^2}{R^2}\right)z_b^2}+\frac{1}{2\exp\left(\displaystyle\frac{3z_a^2}{R^2}\right)z_a^2}+
 \frac{3\,\mbox{\rm Ei}\left(1,\frac{3z_b^2}{R^2}\right)}{2R^2}-\frac{3\,\mbox{\rm Ei}\left(1,\frac{3z_a^2}{R^2}\right)}{2R^2}\right),
\end{equation*}
where $z_a$ depends on  $z_b$ with formulas ~ \eqref{eq27}, \eqref{eq28} and $\mbox{\rm Ei}(1,x)$ is the exponential integral.
The dependence of the relative   area  of the trapped surface is shown graphically at the energies  $E =220$\,GeV and $E =2$\, GeV in Fig. \ref{fig4}. It can be seen that the maximum value of the trapped surface area   $s$ is attained at infinite $z_b$:
\begin{equation*}
s|_{z_b\to\infty}=\frac34\,\frac{{L}^{3}}{G_5{R}^{2}} \left( -\mbox{\rm Ei} \left( 1,{\frac{3z_a^{2}}{{R}^{2}}} \right) +\frac13\,\frac{{R}^{2}}{{{\exp}\left({\displaystyle\frac {3z_a^{2}}{{R}^{2}}}\right)}z_a^{2}} \right),
\end{equation*}
where $z_a$ is given by \eqref{eq29}.
 We note that the expression in the parentheses is always positive.

\begin{figure}[t]
\centering
\includegraphics[height=5cm]{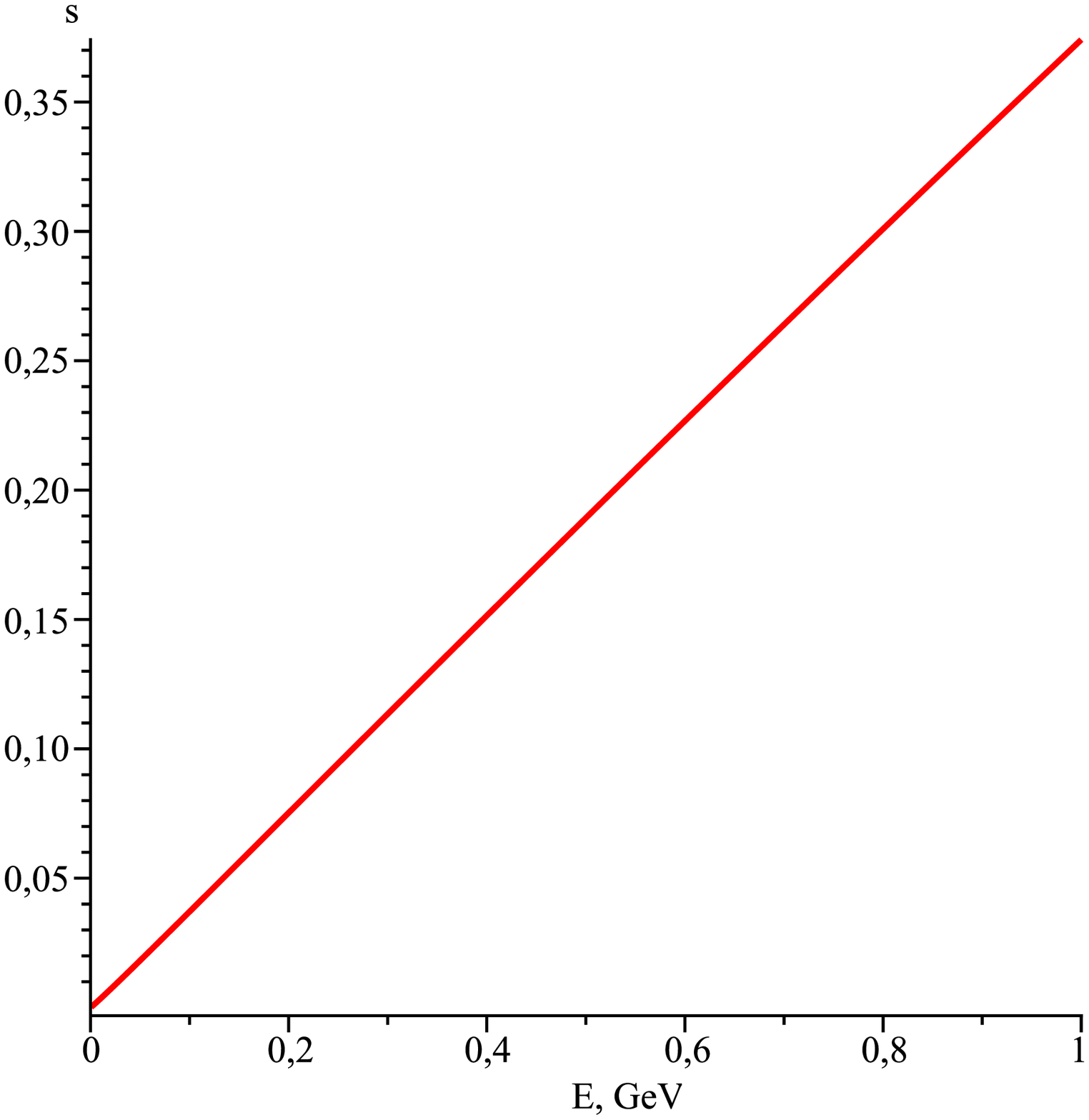}\,A.$\quad$
\includegraphics[height=5cm]{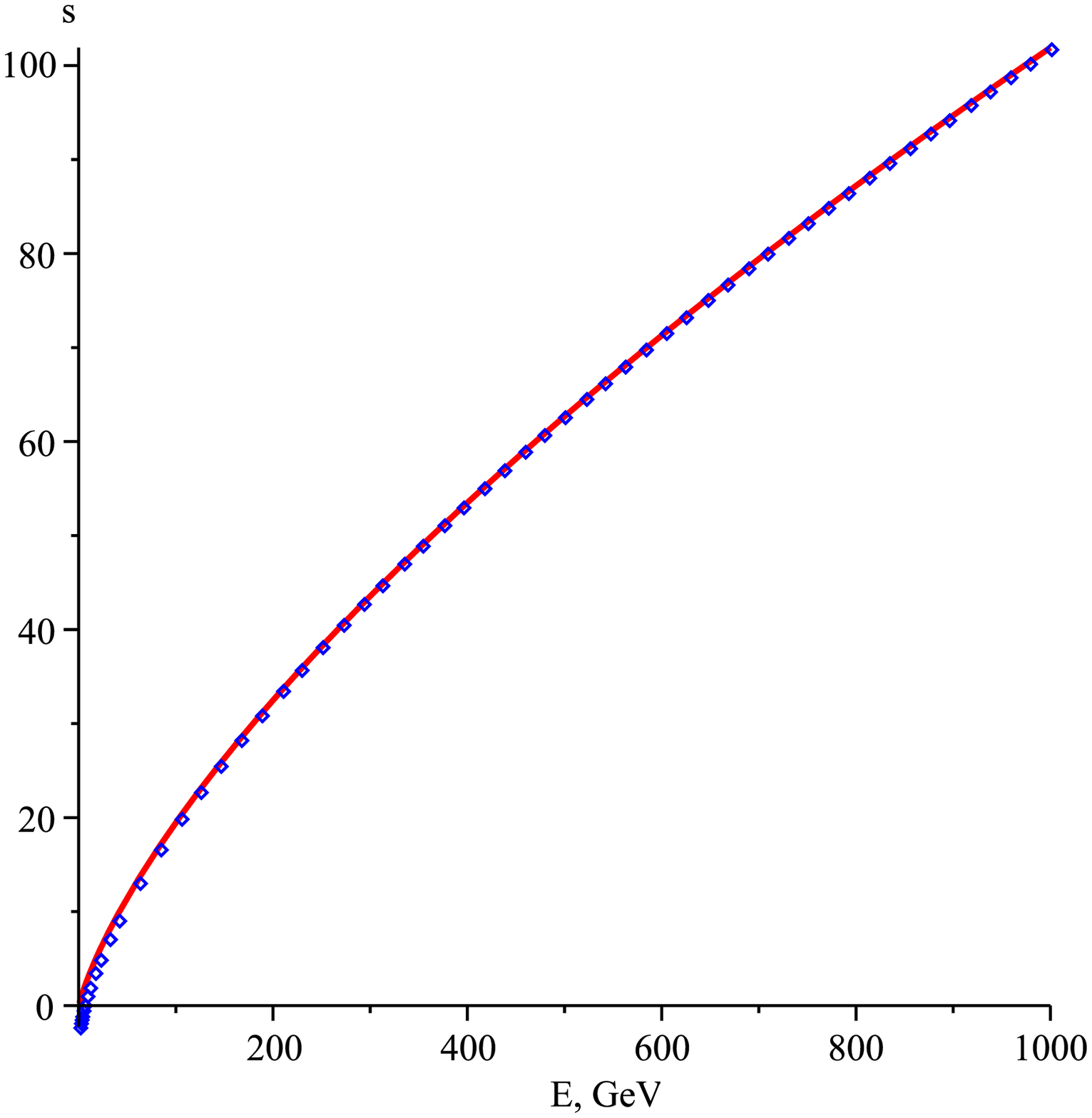}\,B.
\caption{The dependence of the relative area of the trapped surface on energy at low energies  (Fig. A.  at $0<E<1$ GeV ) and at high energies  (Fig. B. at $0<E<1000$ GeV) for $b=(L/z){\mathrm e}^{-z^2/R^2}$. The function approximating
the calculated dependence $E^{2/3}(1 +0.007\ln E) -3$ is
  shown in bold dashes.}\label{fig5}
\end{figure}

Fig. \ref{fig5} shows the dependence of the relative area of the trapped surface on the energy (at law energies in Fig. \ref{fig5} A. and hight energies in Fig. \ref{fig5} B.). Fig. \ref{fig5} B. also shows the function $E^{2/3}(1+0.007\ln E)-3$ approximating the obtained dependence
at  $10\,\text{GeV}\lesssim E<1\,\text{TeV}$.

\section{Mixed  factor of the form $b=\left(\frac{L}{z}\right)^{a}\exp\left(-\frac{z^2}{R^2}\right)$}\label{sec6}

The equation of the domain-wall wave profile in the space with the mixed $b$-factor of the form $b=\left(\displaystyle\frac{L}{z}\right)^{a}\exp\left(-\displaystyle\frac{z^2}{R^2}\right)$ is written as
\be
\label{eq31}
\left(\partial^2_z+\frac{3b^\prime}{b}\partial_z\right)\phi^w(z)=-\frac{16\pi G_5 E}{L^2}\frac{\delta(z-z_*)}{b^3(z)}.\ee
As in the cases considered above, the solution of (\ref{eq31}) is given as
\be
\phi^w(z)=\phi_a\Theta(z_*-z)+\phi_b\Theta(z-z_*),
\ee
where
\bea
&&\phi_a=C_a\int_{z_a}^z b^{-3}dz,\\
&&\phi_b=C_b\int_{z_b}^z b^{-3}dz.
\eea
The constants $C_a$ and $C_b$ can be represented in the form
\bea
&&C_a=\frac{16\pi G_5 E}{L^2}\frac{\int_{z_b}^{z_*} b^{-3}dz}{\int
_{z_b}^{z_a} b^{-3}dz},\label{STAR1}\\
&&C_b=\frac{16\pi G_5 E}{L^2}\frac{\int_{z_a}^{z_*} b^{-3}dz}{\int_{z_b}^{z_a} b^{-3}dz}\label{STAR2}.
\eea
Using the conditions of the  trapped surface formation, we get
\bea
\frac{8\pi G_5 E}{L^2}b^{-3}(z_a)\frac{\int_{z_b}^{z_*} b^{-3}dz}{\int^{z_a}_{z_b} b^{-3}dz}=1,\label{star1}
\eea
\bea
\frac{8\pi G_5 E}{L^2}b^{-3}(z_b)\frac{\int_{z_a}^{z_*} b^{-3}dz}{\int_{z_b}^{z_a} b^{-3}dz}=-1.\label{star2}
\eea
Using the designation $\int_{z_i}^{z_j}b^{-3}dz=F(z_j)-F(z_i)$ and equations \eqref{star1}, \eqref{star2},
we obtain  the relations between points $z_*$, $z_a$, $z_b$ and $z_a$, $z_b$ accordingly

\be F(z_*)=\frac{b^{-3}(z_b)F(z_a)+b^{-3}(z_a)F(z_b)}{b^{-3}(z_a)+b^{-3}(z_b)},\label{int b}\ee

\be b^{-3}(z_a)=\frac{b^{-3}(z_b)}{\displaystyle\frac{8\pi G_5 E}{L^2}b^{-3}(z_b)-1}.\label{b at z_A}\ee

 Relation \eqref{b at z_A} for the factor $b(z)=\left(\displaystyle\frac{L}{z}\right)^a\exp\left(-\displaystyle\frac{z^2}{R^2}\right)$   can be represented in the form
 \be \left(\frac{z_a}{L}\right)^{3a}\exp\left(\frac{3z_a^2}{R^2}\right)=\frac{\left(\frac{z_b}{L}\right)^{3a}\exp\left(\frac{3z_b^2}{R^2}\right)}{\frac{8\pi G_5 E}{L^2} \left(\frac{z_b}{L}\right)^{3a}\exp\left(\frac{3z_b^2}{R^2}\right)-1},\ee
with a solution
 \be \label{lll}
  z_a=\sqrt{\frac{a{R}^{2}}{2}\mbox{\rm W}\left( \frac{2\,{L}^{2}}{a{R}^{2}}{\,{\exp}\left({\frac {2{{z_b}}^{2}}{a{R}^{2}}}\right)} \left(  \frac{\left( {
\frac {{z_b}}{L}} \right) ^{3\,a}{L}^{2}}{ 8\,\pi \,G_5E \left(
{\frac {{z_b}}{L}} \right) ^{3\,a}{{\exp}\left({\frac {3z_b^{
2}}{{R}^{2}}}\right)}-{L}^{2} } \right) ^{\frac{2}{3{a}}}
 \right)},
\ee
one may rewrite (\ref{lll}) in the equivalent form as
\be z_a=R\sqrt{\frac{a}{2}}\sqrt{\mbox{\rm W}\left(\frac{2L^2}{aR^2}\left(\frac{\left(\frac{z_b}{L}\right)^{3a}\exp\left(\frac{3z_b^2}{R^2}\right)}{\frac{8\pi G_5E}{L^2}\left(\frac{z_b}{L}\right)^{3a}\exp\left(\frac{3z_b^2}{R^2}\right)-1}\right)^{\frac{2}{3a}}\right)}.
\ee
This relation has the simplest form at  $a=1/3$, namely
\be
z_a=R\sqrt{\frac{1}{6}}\sqrt{\mbox{\rm W}
\left(\frac{6L^2}{R^2}\left(\frac{\left(\frac{z_b}{L}\right)\exp\left(\frac{3z_b^2}{R^2}\right)}{\frac{8\pi G_5E}{L^2}\left(\frac{z_b}{L}\right)\exp\left(\frac{3z_b^2}{R^2}\right)-1}\right)^{2}\right)}.
\ee
The behavior of $z_a$ is presented at  fixed energies in Fig. \ref{Z_A,S}.
When $z_b\rightarrow \infty,$ we obtain \be\left(\frac{z_a}{L}\right)^{3a}\exp\left(\frac{3z_a^2}{R^2}\right)\rightarrow\frac{L^2}{8\pi G_5 E},\ee
\be z_a|_{z_b\to\infty}=R\sqrt{\frac{a}{2}}\sqrt{\mbox{\rm W}\left(\frac{2L^2}{aR^2}\left(\frac{L^2}{8\pi G_5E}\right)^{\frac{2}{3a}}\right)}.\label{ZA}\ee
Substituting   \eqref{b at z_A} to \eqref{int b}, we obtain the relation
\bea
\label{B}
&&\frac{8\pi G_5 E}{L^2}b^{-3}(z_b)=
\frac{\Gamma\left(\frac{3a+1}{2},-\frac{3z^2_b}{R^2}\right)-\Gamma\left(\frac{3a+1}{2},-\frac{3z^2_a}{R^2}\right)}
{\Gamma\left(\frac{3a+1}{2},-\frac{3z^2_*}{R^2}\right)-\Gamma\left(\frac{3a+1}{2},-\frac{3z^2_a}{R^2}\right)}.
\eea
For the case of $ a=1/3$ it is written as
\bea
&&\frac{8\pi G_5 E}{L^2}b^{-3}(z_b)=\frac{\exp\left( \frac{3zb^2}{R^2}\right)-\exp\left(\frac{3za^2}{R^2}\right)}{\exp\left( \frac{3z_{*}^2}{R^2}\right)-\exp\left(\frac{3za^2}{R^2}\right)}.
\eea
Applying to (\ref{B}) the $\Gamma$ function property
\be\Gamma(A,X)=\Gamma(A)-\frac{X^A\,\,_1F_1(A,A+1,-X)}{A},\label{Gamma}\ee
we have
\bea
&&\frac{8\pi G_5 E}{L^2}b^{-3}(z_b)=
\frac{z_a^{\frac{3a+1}{2}}\,\,_1F_1\left(\frac{3a+1}{2},\frac{3a+3}{2},\frac{3z^2_a}{R^2}\right)
-z_b^{\frac{3a+1}{2}}\,\,_1F_1\left(\frac{3a+1}{2},\frac{3a+3}{2},\frac{3z^2_b}{R^2}\right)}{z_a^{\frac{3a+1}{2}}\,\,_1F_1\left(\frac{3a+1}{2},\frac{3a+3}{2},\frac{3z^2_a}{R^2}\right)-z_*^{\frac{3a+1}{2}}\,\,_1F_1\left(\frac{3a+1}{2},\frac{3a+3}{2},\frac{3z^2_*}{R^2}\right)}.
\eea
With the help of the relation
\be
\,\,_1F_1(\alpha,\gamma,z)=\sum_{k=0}^\infty\frac{(\alpha)_k}{(\gamma)_k}\frac{z^k}{k!},
\ee
where $_1F_1(\alpha,\gamma,z)$ is the confluent hypergeometric function,  $ (\alpha)_k $ and $(\gamma)_k$ are the pochhammer symbols,
we have
\be \label{53}\frac{8\pi G_5 E}{L^2} \left(\frac{z_b}{L}\right)^{3a}\exp\left(\frac{3z_b^2}{R^2}\right)=
\sum_{k=0}^\infty\frac{\left(\frac{3z_a^2}{R^2}\right)^{k+\frac{3a+1}{2}}-\left(\frac{3z_b^2}{R^2}\right)^{k+\frac{3a+1}{2}}}{\left(\frac{3z_a^2}{R^2}\right)^
{k+\frac{3a+1}{2}}-\left(\frac{3z_*^2}{R^2}\right)^{k+\frac{3a+1}{2}}}.\ee
 Accordingly to (\ref{53}) we observe that $z_*\rightarrow z_a$ from the right side, if $z_b\rightarrow\infty.$

For an arbitrary $a$ the relation \eqref{int b} can be represented in the form:
$$
z_*^{2a}{}_1F_1\left(\frac{3a+1}{2};\frac{3(a+1)}{2};\frac{3z_*^2}{R^2}\right)=
$$

$$
=\frac{z_a^{2a}z_b^{2a}\left(z_b^a e^{\left(\frac{3z_b^2}{R^2}\right)}{}_1F_1\left(\frac{3a+1}{2};\frac{3(a+1)}{2};\frac{3z_a^2}{R^2}\right)
+z_a^ae^{\left(\frac{3z_a^2}{R^2}\right)}{}_1F_1\left(\frac{3a+1}{2};\frac{3(a+1)}{2};\frac{3z_b^2}{R^2}\right)\right)}
{z_a^{3a}\exp\left(\frac{3z_a^2}{R^2}\right)+z_b^{3a}\exp\left(\frac{3z_B^2}{R^2}\right)}.
$$
This equation is rather difficult for analytic solutions.
However, in the case $a=1/3$ the expression \eqref{int b}
can be represented  as the following one
\begin{equation}
\exp\left(\frac{3z_*^2}{R^2}\right)=\frac{z_a+z_b}{z_a\exp\left(-\frac{3z_b^2}{R^2}\right)+z_b\exp\left(-\frac{3z_a^2}{R^2}\right)},
\end{equation}
which
 has the analytical  solution
\begin{displaymath}
z_{*}=\sqrt{\frac{R^2}{{3}}\ln\biggm(\cfrac{z_a+z_b}{z_a\exp(-\frac{3z_b^2}{R^2})+z_b\exp(-\frac{3z_a^2}{R^2})}\biggm)}.
\end{displaymath}
The functions $\phi_a,$ $\phi_b$ can be represented such as
\begin{equation}\label{phi_a}
    \phi_a=\frac{16\pi G_5 E}{L^2}\cdot\frac{\int_{z_b}^{z_*}b^{-3}dz\cdot\int_{z_a}^{z}b^{-3}dz}{\int_{z_b}^{z_a}b^{-3}dz},
\end{equation}
\begin{equation}\label{phi_b}
    \phi_b=\frac{16\pi G_5 E}{L^2}\cdot\frac{\int_{z_a}^{z_*}b^{-3}dz\cdot\int_{z_b}^{z}b^{-3}dz}{\int_{z_b}^{z_a}b^{-3}dz}.
\end{equation}
Considering the integral
\begin{displaymath}
\int b^{-3}dz=\left(\frac{3^{-\frac12-\frac{3a}{2}}}{2}\right)\left(-\frac{z^2}{R^2}\right)^{-\frac12-\frac{3a}{2}}\left(\frac{L}{z}\right)^{-3a}z
\left(\Gamma\left(\frac{3a}{2}+\frac12\right)-\Gamma\left(\frac{3a}{2}+\frac12,-\frac{3z^2}{R^2}\right)\right)+C,
\end{displaymath}
and using the property \eqref{Gamma}, one obtains
\begin{equation}
\int b^{-3}dz=\frac{z\left(\frac{L}{z}\right)^{-3a}{}_1F_1\left(\frac{3a+1}{2},\frac{3(a+1)}{2},\frac{3z^2}{R^2}\right)}{3a+1}+C.
\end{equation}
Introducing  new designation
\begin{equation}\label{ups}
  \Upsilon(z)= z\left(\frac{L}{z}\right)^{-3a}{}_1F_1\left( \frac{3a+1}{2}, \frac{3(a+1)}{2}, \frac{3z^2}{R^2}\right),
\end{equation}
So the expressions
\eqref{phi_a} and \eqref{phi_b} are represented in the form
\begin{eqnarray}\label{phia}
\phi_a=\frac{16 \pi G_{5} E}{(3a+1)L^2}\frac{\biggm( \Upsilon(z_b)-\Upsilon(z_*) \biggm)\biggm( \Upsilon(z)-\Upsilon(z_a) \biggm)}{\Upsilon(z_b)-\Upsilon(z_a)},
\end{eqnarray}
\begin{eqnarray}\label{phib}
\phi_b=\frac{16 \pi G_{5} E}{(3a+1)L^2}
\frac{\biggm( \Upsilon(z_*)-\Upsilon(z_a) \biggm)\biggm( \Upsilon(z_b)-\Upsilon(z) \biggm)}{\Upsilon(z_b)-\Upsilon(z_a)}.
\end{eqnarray}
The expression (\ref{ups})  has the following form at $ a=1/3$
\be\label{profn=1/3}\Upsilon(z)=\frac{R^2}{3L}\biggm( \exp\left(\frac{3z^2}{R^2}\right)-1 \biggm),  \ee
\begin{figure}[t] \centering
\includegraphics[height=7cm]{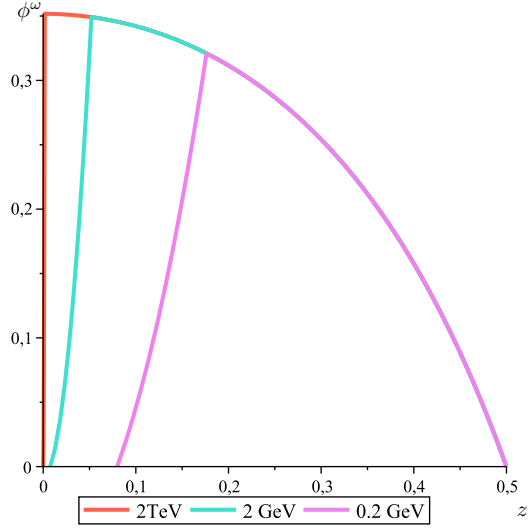}
\caption{The profile $\phi^{\omega}$ corresponding to  $b(z)=\left(\frac{L}{z}\right)^a\exp\left(-\frac{z^2}{R^2}\right)$ at $z_b=0.5$ fm, $a={1}/{3}$, $R=1$ fm, $E=0.2$ GeV, $E=2$ GeV, $E=2$ TeV. }\label{lolo}
\end{figure}
so the functions $\phi_a $ and $\phi_b$ are represented as
\begin{equation}\label{phia1}
  \phi_a=\frac{8}{3}\frac{\pi G_{5} E R^2}{L^3}\frac{\biggm(\exp\left(  \frac{3z_{b}^{2}}{R^2}\right)- \exp\left(\frac{3z_{*}^2}{R^2} \right)\biggm)
  \biggm(\exp\left(  \frac{3z^2}{R^2}\right) - \exp\left(\frac{3z_{a}^2}{R^2} \right)\biggm)}{\exp\left(\frac{3z_{b}^2}{R^2}\right)- \exp\left(\frac{3z_{a}^2}{R^2}\right)},
\end{equation}
\begin{equation}\label{phib1}
  \phi_b=\frac{8}{3}\frac{\pi G_{5} E R^2}{L^3}\frac{\biggm(\exp\left(  \frac{3z_{*}^{2}}{R^2}\right)- \exp\left(\frac{3z_{a}^2}{R^2} \right)\biggm)
  \biggm(\exp\left(  \frac{3z_{b}^2}{R^2}\right) - \exp\left(\frac{3z^2}{R^2} \right)\biggm)}{\exp\left(\frac{3z_{b}^2}{R^2}\right)- \exp\left(\frac{3z_{a}^2}{R^2}\right)}.
\end{equation}
The given profiles at $a=\displaystyle{1}/{3}$, $G_5=L^3/1.9$, $L=4.4$ fm, $z_b=0.5$ fm are presented in Fig. (\ref{lolo}).

\begin{figure}[t] \centering
\includegraphics[height=7 cm]{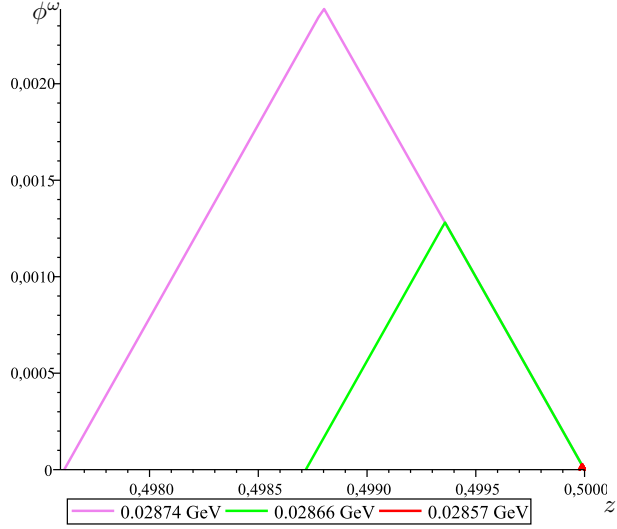}
\caption{The profiles $\phi^{\omega}$ corresponding to  $b(z)=\left(\frac{L}{z}\right)^a\exp\left(-\frac{z^2}{R^2}\right)$ at $z_b=0.5$ fm, $a={1}/{3}$, $R=1$ fm, $E=28.74$ MeV, $E=28.66$ MeV, $E=28.57$ MeV. }\label{pepe}
\end{figure}

Similar to the case discussed in the sections above, due to the condition $z_a<z_b$ we find the range of the energies
 at which black hole is created
\begin{displaymath}
  E>\frac{L^3}{4\pi G_{5}z_{b}\exp\left(\frac{3z_b^2}{R^2}\right)}.
\end{displaymath}
Black hole is not created created as the energy approaches to the value
\begin{equation}\label{energy escape}
  E=\frac{L^3}{4\pi G_{5}z_{b}\exp\left(\frac{3z_b^2}{R^2}\right)},
\end{equation}
 which is  equal to $28.568$ MeV for the parameters $G_{5} = L^{3}/1.9$, $L = 4.4$ fm, $a={1}/{3}$  and $z_{b}=0.5$ fm., (see Fig. \ref{pepe}).

In considering case the relative trapped surface area is $$s=
\left.\frac{\left( {\frac {L}{z}} \right) ^{3a}z\,{{\exp}\left(-{\frac {3{z}^{2}}{2{R}^{2}}}\right)} \left( 2 \left(\frac{3z^2}{R^2}\right)^{\frac{3a-1}{4}}
{{\rm \bf M}\left(\frac{-3a+1}{4},\frac{3(-a+1)}{4},\,{\frac {3{z}^{2}}{{R}^{2}}}\right)}
+3(1-a){{\exp}\left(-{\frac {3{z}^{2}}{2{R}^{2}}}\right)} \right)}{ 2 G_5 \cdot 3 \left(3\,a-1 \right) \left( a-1 \right)}\right|_{z_a}^{z_b},
$$
where ${{\rm\bf M}(\mu, \nu, z)}= \exp\left(-\frac{z}{2}\right) z^{\frac12+\nu}\,_1F_1(\frac12+\nu-\mu, 1+2\nu, z)$  is the  Whittaker function, $a\neq1/3,$ $a\neq1$.

\begin{figure}[t] \centering
\includegraphics[height=5cm]{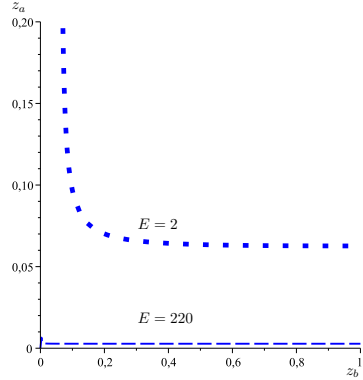}
\caption{ The dependence  of  $z_a$ on $z_b$   corresponding to $a={1}/{2}$  at energies  $2$ GeV and $220$ GeV.
}\label{Z_A,S}
\end{figure}

\begin{figure}[t] \centering
\includegraphics[height=5cm]{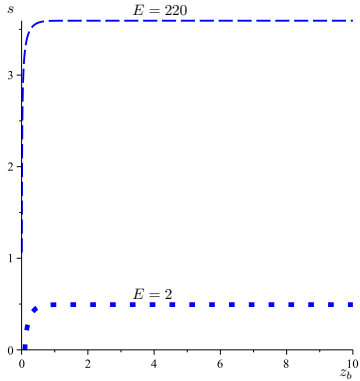}
\caption{ The  dependence of the  relative area  of the trapped surface on $z_b$  corresponding to $a={1}/{2}$  at energies  $2$ GeV and $220$ GeV.
}\label{Z_A,Sb}
\end{figure}

\begin{figure}[t] \centering
\includegraphics[height=5cm]{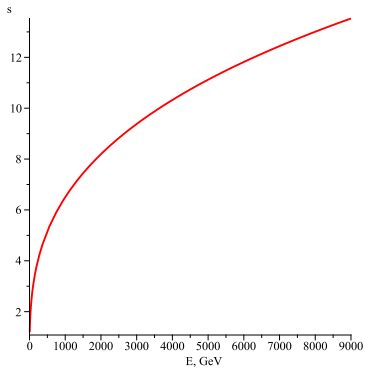}
\caption{The dependence of the relative area of the trapped surface on energy corresponding to $a={1}/{2}$.
}\label{SmixtCase}
\end{figure}

We  graphically represent $s$ as a function of $z_b$  in Fig. \ref{Z_A,Sb} at energies $2$ GeV and $220$ GeV.
Hence,  it is clear that the relative trapped surface area tends to its maximum value at infinite
$z_b$:
\begin{displaymath}
s\rightarrow \frac{\left( {\frac {L}{z_a}} \right) ^{3a}z_a\,{{\exp}\left(-{\frac {3{z_a
}^{2}}{2{R}^{2}}}\right)} \left( 2 \left(\frac{3z_a^2}{R^2}\right)^{\frac{3a-1}{4}}
{{\rm \bf M}\left(\frac{-3a+1}{4},\frac{3(-a+1)}{4},\,{\frac {3{z_a}^{2}}{{R}^{2}}}\right)}
+3(1-a){{\exp}\left(-{\frac {3{z_a}^{2}}{2{R}^{2}}}\right)} \right)}{ 6 G_5 \cdot \left(3\,a-1 \right) \left(1-a \right)},
\end{displaymath}
where $a>\frac13,$ $a\neq 1,$  $z_a$ is defined by \eqref{ZA}.

 For the case corresponding to $a=1/3$ the relative trapped surface area is written as
\be
s= -\frac{L}{4G_5}\textmd{Ei}{\left( 1,\frac{3z^2}{R^2}\right)}\Biggm|_{z_a}^{z_b},
\ee
and
\be
s\rightarrow \frac{L}{4G_5}\textmd{Ei}\left( 1,\frac{3z_{a}^2}{R^2}\right)
\ee
at $ z_b\rightarrow \infty$.

\begin{figure}[t] \centering
\includegraphics[height=4.9cm]{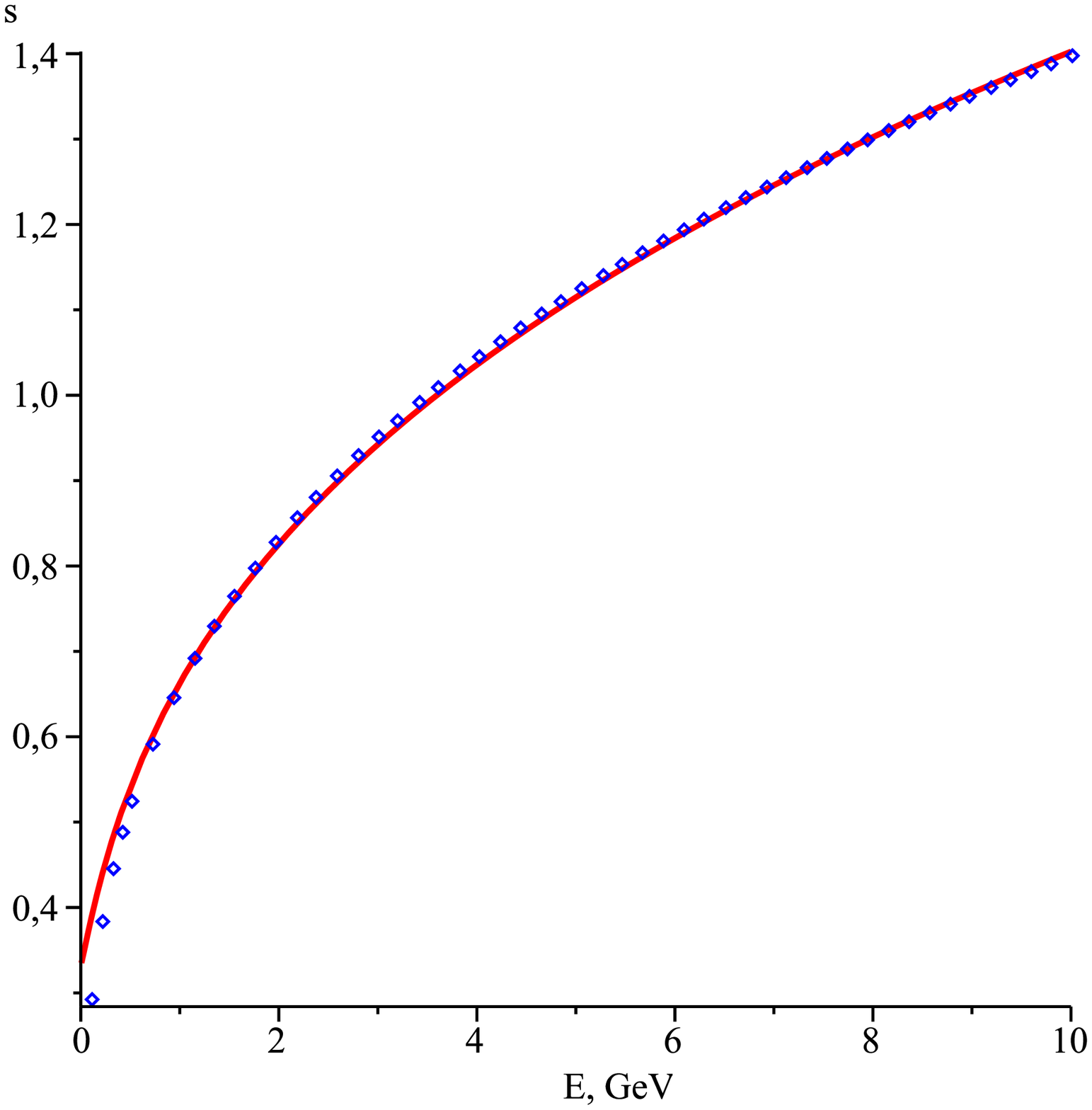} A.\includegraphics[height=4.9cm]{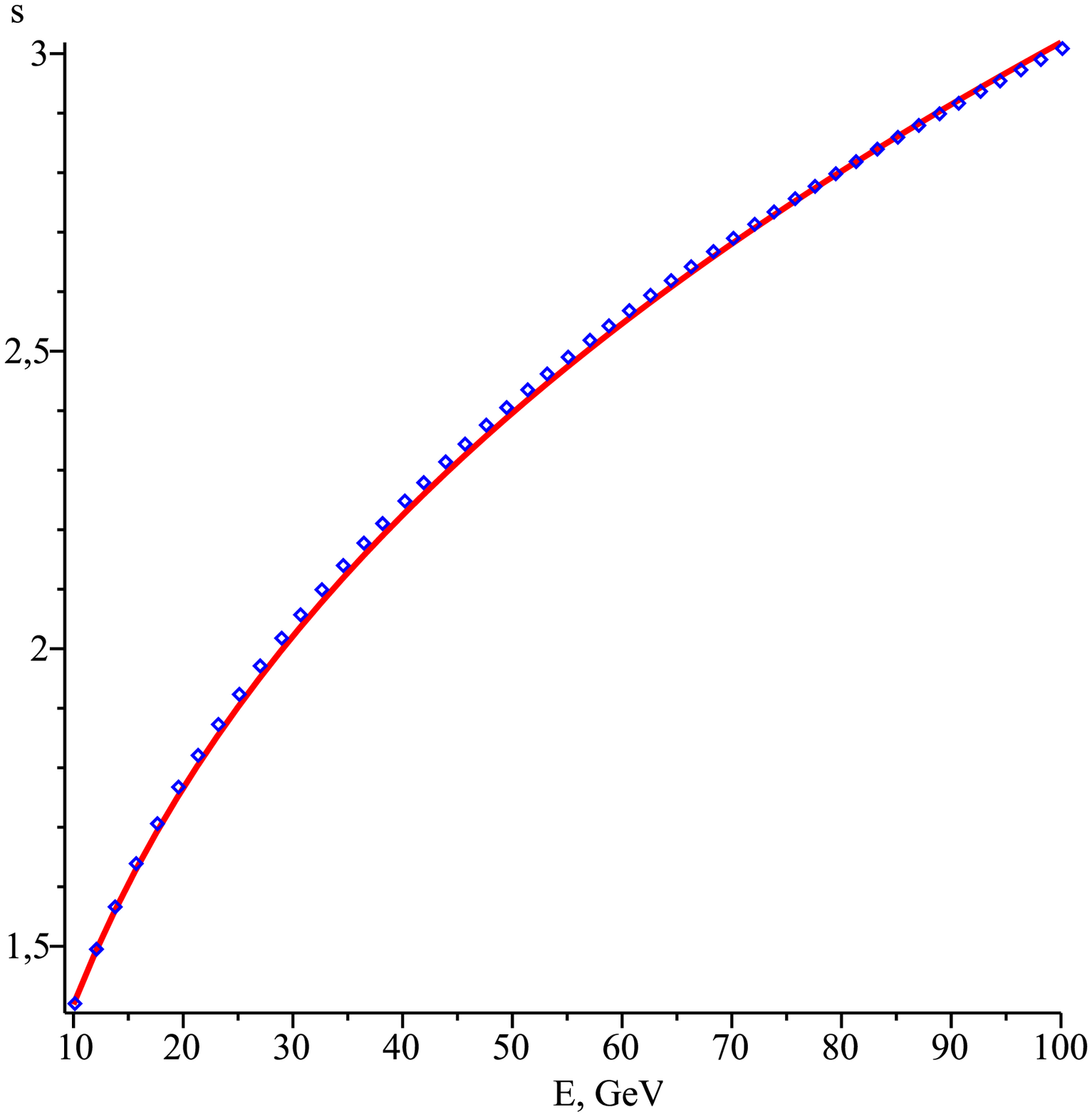}  B.\includegraphics[height=4.9cm]{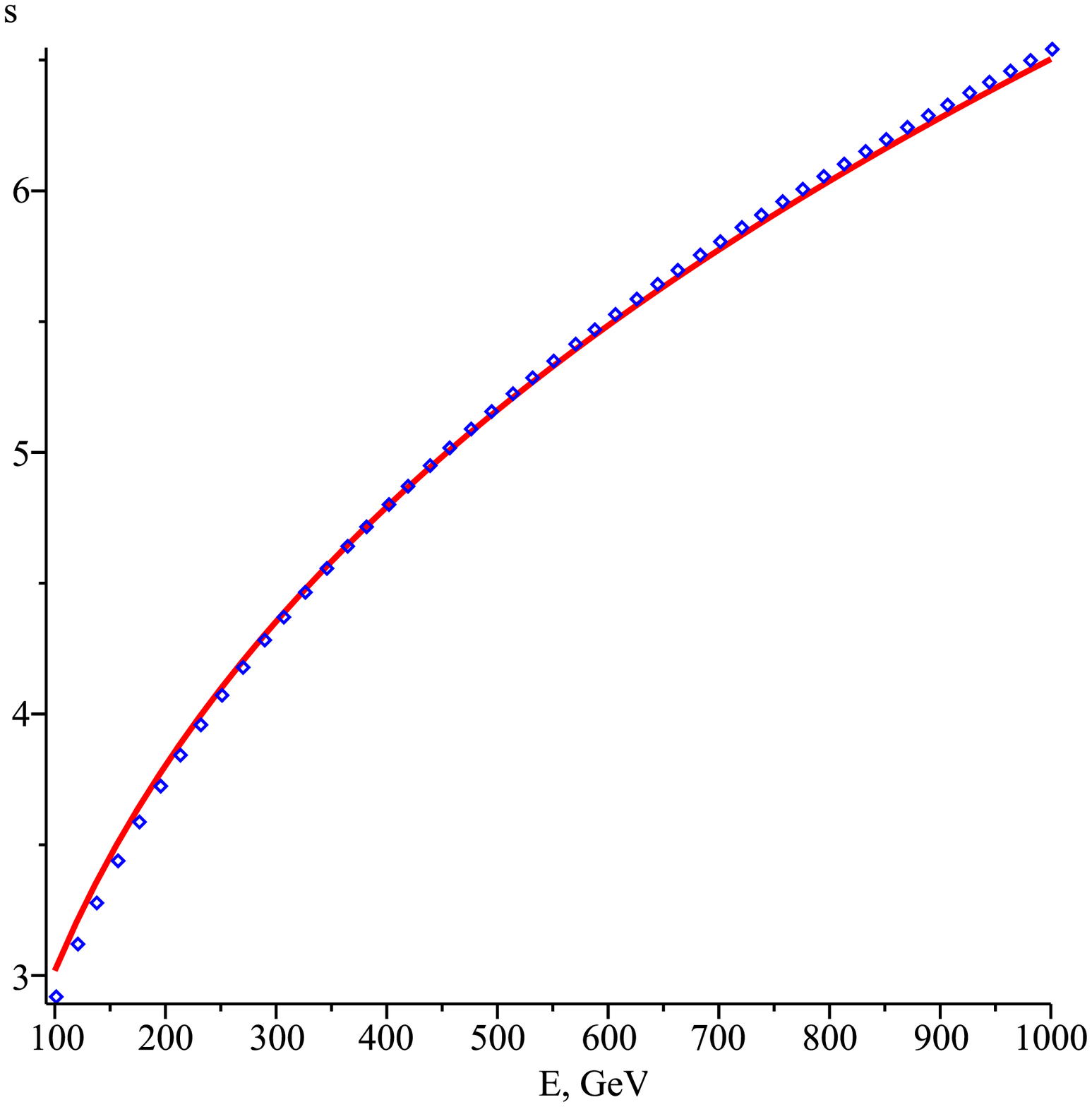} C.
\caption{ The dependence of the relative area of the trapped surface on energy is shown in red line and the function approximating the calculated dependence corresponding to  $a=1/2$ is shown in blue points  (Fig. A. $ \frac{E^{0.3}\left(57-29.75(\ln(E+100))\right)-7}{2 G_5}$  at  $0<E<10$ CeV ),
 (Fig. B. $\frac{E^{0.3}\left(61-45.05(\ln(E+100))\right)-24}{2 G_5}$  at $10<E<100$ GeV)
 and  (Fig. C. $\frac{E^{0.3}(81-5.95\ln(E+100))-67}{2G_5}$  at  $10^2<E<10^3$ GeV).}\label{Ap2}
\end{figure}

\newpage

\section{Conclusion}\label{sec4}
We have investigated the possibility of the black hole formation in the domain-wall collisions  in modified  $\mathrm{AdS}_5$  spaces with  $b$-factors.
We considered several types of  $b$-factor: power-law, exponential and mixed. We analyzed the dependence of entropy on the energy of colliding ions
 in the spaces with  $b$-factors based on the analysis of the conditions for forming the trapped surfaces.
 With the AdS/CFT duality taken into  account, the obtained results allow modeling the dependence of  multiplicity  of the  produced particles  on the energy of the colliding heavy-ions. We note that the results for the power-law factors are agree  with the conclusion in
 the  previously examined cases of central collisions of pointlike sources.  The exponential factors of the collision domains do not lead to additional logarithms which take place in the case of central collisions of point sources  \cite{12} in the presence  of the exponential $b$-factors;
 nevertheless, additional logarithms appear when the mixed factors are considered.
 The derived results can be used to  compare with the experimental curves for the multiplicity of  particle formation in heavy-ion collisions.

\subsection*{Acknowledgments}
The results introduced in this paper were partially presented by one of the authors (I. Ya. A.) at the Fourth International Conference "Models in Quantum Field Theory" (MQFT-2012) devoted to the memory of A. N. Vassiliev.

This work was partially supported by the Russian Foundation for Basic Research (grants 11-01-00894-a (I. A., E. P.) and 1202-31109-mol\_a (E. P., T. P.))
and  by the Russian Ministry of Education and Science under grant NSh-3042.2014.2 (E.P.).

\end{document}